\def\elnino{El Ni{\~n}o~}

\def\degree{\hbox{$^\circ$}}

\documentclass[review]{elsarticle}

\usepackage{amssymb}

\usepackage{graphicx}
\usepackage{hyperref}
\usepackage{amssymb,amsmath}
\usepackage{makecell}

\usepackage{lineno}

\journal{Chaos, Solitons \& Fractals}

\begin{document}

\begin{frontmatter}



\title{Earthquake activity as captured using the network approach}


\author[BIDR]{Yosef Ashkenazy\corref{cor1}}
\ead{ashkena@bgu.ac.il}
\cortext[cor1]{corresponding author}
\author[GSI]{Ittai Kurzon}
\author[BIU]{Eitan Asher}
\address[BIDR]{Department of Environmental Physics, BIDR,
Ben-Gurion University, Midreshet Ben-Gurion, Israel}
\address[GSI]{Geological Survey of Israel, Jerusalem, Israel}
\address[BIU]{Physics Department, Bar-Ilan University, Ramat-Gan, Israel}

\begin{abstract}
Earthquakes are a major threat to nations worldwide. Earthquake detection and forecasting are important and timely scientific challenges, not only for their obvious social impacts, but also because they reflect the actual degree of understanding of the physical processes controlling seismic event occurrences. Here, we propose an alternative approach for evaluating and understanding the dynamics of seismic events. The approach is based on the phase between the waveform signals of many stations, enabling detecting the evolution of relatively small magnitudes, down to Mw 1.3. We constructed a time-evolving network in which the network nodes are the stations, while the links are the level of correspondence between the stations' signals. The links' weights are quantified using the following statistical methods: cross-correlation, synchronization, mutual information, and coherence. Each of these methods reflects a different aspect of the phase relations between the waveforms of different stations in a given time window. We then developed global measures to study the properties of the time-evolving network of seismic activity. The global measures include the leading eigenvalues of the network links, the number of links above a certain threshold, and $k$-means clustering. We show that the network and its corresponding global measures vary significantly during seismic events. The results are based on detailed waveform station data and detailed catalogs from Southern California; our analysis focused on 27 mainshocks, during which we examined one-day data prior to the occurrence of the mainshock, as well as one hour of data following it. Among all the measures we investigated, we found that the coherence measure using the $k$-means clustering procedure exhibits the best performance. This technique correctly identifies earthquake events with magnitudes larger than 2.5 and exhibits moderate performance for weaker earthquakes with magnitudes larger than 1.3. 
\end{abstract}


\begin{highlights}
\item We analyzed waveform data from the California region. 
\item We constructed a time-evolving network of different statistical measures.
\item The coherence measure using $k$-means clustering exhibits peaks that match earthquake events larger than 2.5. 
\end{highlights}


\end{frontmatter}


\section{Introduction}
\label{sec:intro}

There are numerous examples of the destructive power of major earthquakes, and thus, it is not surprising that intensive efforts have been devoted to coping with these events. Building standards and social strategies have been adopted to minimize earthquake damage, especially in tectonically active regions \cite[e.g.,][]{Freddi-Galasso-Cremen-et-al-2021:innovations}. In addition, Earthquake Early Warning Systems (EEWSs) are used to alert the population and provide advanced warnings in areas that are located at a sufficient distance from the earthquake's epicenter \cite[e.g.,][]{ Kuyuk-Allen-2013:optimal}. Such early alarms can be used to shutdown sensitive facilities including (nuclear or gas) power plants, elevators, and more. EEWSs quickly recognize the initial, less destructive, but faster, primary $p$ (pressure) wave to warn of the arrival of the secondary and more destructive secondary $s$ (shear) wave. EEWSs utilize their significantly higher communication speed, in comparison to the slower speed of seismic waves (order of several km/s), to acquire and process real-time data. Depending on the distance from the earthquake’s epicenter, the warning time is on the order of seconds to tens of seconds \cite[][]{Gasparini-Manfredi-Zschau-2007:earthquake}. 

An alternative approach, or an additional layer, lies in forecasting earthquakes. Previous studies \cite[e.g.,][]{Jordan-Chen-Gasparini-et-al-2011:operational} have suggested different earthquake precursors; however, these attempts have attained only limited success. Another more successful approach is the use of statistical models, such as the Epidemic-Type Aftershock Sequence (ETAS) model \cite[][]{Ogata-1988:statistical}, to statistically forecast the aftershocks after mainshocks. Such models are built on empirical earthquake laws: the Gutenberg-Richter law \cite[][]{Gutenberg-Richter-1944:frequency, Gutenberg-Richter-1956:earthquake} and the Omori-Utsu law \cite[][]{Omori-1894:after, Utsu-1961:statistical}. These models, however, are not capable of forecasting mainshocks \cite[][]{Ogata-1988:statistical, Ogata-1998:space,Gerstenberger-Wiemer-Jones-et-al-2005:real}, and their aftershock prediction success is limited \cite[][]{Taroni-Marzocchi-Schorlemmer-et-al-2018:prospective,Woessner-Hainzl-Marzocchi-et-al-2011:retrospective}. It is clear that even a slight improvement in the skill of forecasting major earthquakes and aftershocks is of great importance. In addition, more detailed earthquake catalogs \cite[e.g.,][]{Ross-Trugman-Hauksson-et-al-2019:searching, Tan-Waldhauser-Ellsworth-et-al-2021:machine}, together with high-resolution waveform data from sufficiently close stations, may improve the understanding of earthquake forecasting and, hopefully, the capabilities themselves. We are not attempting to provide a better forecasting scheme or EEWS, but to suggest an alternative approach to analyzing and characterizing evolving earthquakes. More specifically, we suggest waveform processing, using the whole network simultaneously, as opposed to the common practice of processing each station independently or using multiple stations in order to detect the event's epicenter and the direction of wave propagation. The network's links are based on measures that quantify the level of correspondence between the waveform signals of different stations, ignoring, in the first stage, the amplitude of the signals and focusing on the relative phase between them. 

Networks consist of nodes and the links that connect the nodes \cite[][]{Cohen-Havlin-2010:complex}. A link can be defined via the cross-correlation function \cite[e.g.,][]{Tsonis-Roebber-2004:architecture, Zhang-Fan-Chen-et-al-2019:significant} and, to a lesser degree, through the level of synchronization \cite[e.g.,][]{Zhang-Hu-Kurths-et-al-2013:explosive}; other measures, such as mutual information and coherence, can be used as well. In seismology, the most common use of cross-correlation is for post-processing seismic waveforms, in which it is used to locate additional seismic events that were not originally identified in the real-time processed data \cite[e.g.,][]{Schaff-Waldhauser-2010:one,Gao-Kao-2020:optimization}. In addition, cross-correlation techniques have been used to refine the arrival of different seismic phases and to ``time-shift'' them, leading to improved identification of seismic event epicenters and a reduction in the associated uncertainties \cite[e.g.,][]{Hauksson-Shearer-2005:southern, Shearer-Hauksson-Lin-2005:southern}. In recent years, cross-correlation in time domains and frequency domains (using coherence) has been established as a very efficient tool for seismic interferometry \cite[e.g.,][and references therein]{Snieder-Miyazawa-Slob-et-al-2009:comparison, Wapenaar-Van-Ruigrok-et-al-2011:seismic}, seismic noise tomography \cite[][]{Roux-Sabra-Gerstoft-et-al-2005:p, Shapiro-Campillo-Stehly-et-al-2005:high}, the analysis of mini-seismic arrays \cite[][]{Cansi-Plantet-Massinon-1993:earthquake, Eisermann-Ziv-Wust-Bloch-2018:array}, and beamforming \cite[][]{Ruigrok-Gibbons-Wapenaar-2017:cross}.

In this study, we used available seismic station data (waveform) to construct a network of earthquake activity. The network nodes are the station locations, while the links between these nodes (stations) are defined by the statistical relations between them, e.g., by the level of lagged cross-correlation between the stations; we also used other time and frequency domain measures including synchronization, mutual information, and coherence \cite[see, e.g.,][]{Shi-Seydoux-Poli-2021:unsupervised}. We defined the network’s collective measures and studied the continuous temporal evolution of these measures. Our results indicate that the levels of network correlation, synchronization, mutual information, and coherence increase (or decrease) during seismic events; these changes can be used for real-time detection of seismic events. Moreover, we used the $k$-means clustering to analyze the clusters associated with the different measures and also to analyze all the measures simultaneously using the time-evolving network; the required computation time is not long, such that the suggested procedure can be applied, in principle, in real time.

The various advanced signal processing tools we use here have been used in the past to study seismic data: a) the synchronization technique has been applied to study the global seismic catalog \cite[e.g.,][]{Bendick-Mencin-2020:evidence}; b) mutual information has been used to visualize and analyze seismic data \cite[e.g.,][]{Machado-Lopes-2013:analysis, Ramirez-Rojas-Paez-Hernandez-Luevano-2013:cross}; c) the coherence technique has been widely applied in seismic array processing, though in ways differing from our study, mainly in the aperture scale of the examined seismic network \cite[e.g.,][]{Song-Pitarka-Somerville-2009:exploring,Ding-Trifunac-Todorovska-et-al-2015:coherence}; and d) the $k$-means clustering has been applied to investigate earthquake catalogs \cite[][]{Ansari-Noorzad-Zafarani-2009:clustering} and to classify regions of seismic activity \cite[][]{Weatherill-Burton-2009:delineation, Novianti-Setyorini-Rafflesia-2017:k}. The time-evolving, real-time network approach we propose here implements all the above techniques, processing many stations simultaneously. 

Network analyses \cite[][]{Cohen-Havlin-2010:complex, Tenenbaum-Havlin-Stanley-2012:earthquake} have led to advances in many branches of science where patterns of feedbacks and interactions are taken into account. The conventional analysis of climate records has been broadened using network-based approaches; examples of climate network approach applications include, e.g., the \elnino phenomenon \cite[][]{Gozolchiani-Havlin-Yamasaki-2011:emergence, Ludescher-Gozolchiani-Bogachev-et-al-2013:improved, Fan-Meng-Ashkenazy-et-al-2017:network} and atmospheric Rossby waves \cite[][]{Wang-Gozolchiani-Ashkenazy-et-al-2013:dominant}. To apply a network analysis to a time series of events in two locations, it is necessary to have an inclusion criterion $I(i,j)$. When $I(i,j)$ is met, $i$ and $j$ nodes are defined as connected. $I(i,j)$ is usually based on measures of similarity $W(i,j)$ (a link weight) between records measured in different locations, $i$ and $j$, such as cross-correlation, synchronization, mutual information, and coherence. $I(i,j)$ is met when $W(i,j)$ exceeds (positively or negatively) a certain value. For typical time series in extreme event problems, such as seismic records, there are a few rare spikes on top of a background signal. In such cases, the method of event synchronization \cite[][]{Boers-Bookhagen-Marwan-et-al-2013:complex, Quiroga-Kreuz-Grassberger-2002:event} was found useful.

A complex network can be constructed based on earthquake catalogs in various ways \cite[e.g.,][]{Abe-Suzuki-2003:law, Baiesi-Paczuski-2004:scale, Davidsen-Grassberger-Paczuski-2006:earthquake, Lotfi-Darooneh-2012:earthquakes, Tenenbaum-Havlin-Stanley-2012:earthquake, Rezaei-Darooneh-Lotfi-et-al-2017:earthquakes, Celikoglu-2020:earthquake, He-Wang-Liu-et-al-2020:similar, He-Wang-Zhu-et-al-2021:statistical}. For example, Abe and Suzuki \cite{Abe-Suzuki-2003:law} divided the area of interest into geographical grid locations (nodes), and consecutive earthquakes defined the links; the significance of the links was not considered. This network approach resulted in a scale-free (small world) network in which the probabilities of the node degrees decay as a power law, in which the mainshocks are the network hubs. Tenenbaum {\it et al.} \cite{Tenenbaum-Havlin-Stanley-2012:earthquake} regarded the geographical locations as nodes, and the links between the nodes were computed using the similarity between the activity of the two nodes. The above studies were based on earthquake catalogs for which the time, magnitude, and coordinates of the events are known. Here, however, we apply the network approach to waveform data, aiming to detect seismic events with unknown magnitudes; this is based on the relative phase between the signals of the different stations, but not on the amplitude. The information regarding the amplitude may supplement the information derived from the other measures.


\section{Data and Methods}

\subsection{Data}

\paragraph{Pre-processing:}
\label{sec:per-processing} 
The analysis we employ here is based on waveform data from stations throughout the state of California and its close surroundings. We considered 27 major earthquakes (mainshocks) with magnitudes of $M_w\ge 4.5$ in this region (114\degree{W}-124\degree{W}, 32\degree{N}-42\degree{N}) between 2010 and 2020; see Fig. \ref{fig:map_all_events} for the location of these events and Table \ref{table:events} for information regarding these events. Then, we generated a waveform dataset, using the vertical channel of the velocity waveforms from stations that fall within a radius of 5 degrees around the epicenter of each event. The data was downloaded from the USGS FDSN data server (e.g., \url{https://earthquake.usgs.gov/fdsnws/event/1/}) using the FetchData command. Stations with large data gaps, extremely noisy data, or many non-seismic outliers were filtered out. We imposed a minimum distance between stations of 700 m, to avoid the situation of dense seismic arrays that may bias the network topology. Following the above procedure and based on our results, we typically analyzed $\sim$100 stations’ waveform data per event. For each mainshock, we processed 24 hours of data before the mainshock and one hour after. As the different stations have different recording frequencies and to reduce the computation time, we down-sampled the data to 40 samples per second.

For validation purposes, we used two earthquake catalogs from the California region. The first was constructed from the USGS FDSN data server, covering the region of 114\degree{W}-124\degree{W} and 32\degree{N}-42\degree{N}. We considered events with magnitudes larger than 2.5 as the catalog is complete above this magnitude. We also used the Ross {\it et al.} \cite{Ross-Trugman-Hauksson-et-al-2019:searching} earthquake catalog. This catalog spans the Southern California region (i.e., 114\degree{W}-122\degree{W} and 32\degree{N}-37\degree{N}), and covers only part of the spatial extent of the mainshocks we considered here. The catalog contains events from the beginning of 2008 till the end of 2017, while we analyzed seismic data between 2010 and 2020; thus, part of the data we analyzed here cannot be compared with the Ross {\it et al.} catalog. We considered events with magnitudes larger than 1.3 as the Ross {\it et al.} catalog is complete above this magnitude. 

\subsection{Methods}

\paragraph{Cross-correlation:}
\label{sec:cc}
Given two time series, $x_1(t)$, $x_2(t)$, the lagged cross-correlation function is defined as
\begin{equation}
  C_{x_1,x_2}(\tau)=\frac{E[(x_1(t)-m_{x_1})(x_2(t+\tau)-m_{x_2})]}{s_{x_1}s_{x_2}},
\end{equation}
where $E[\cdot]$ indicates the expectation value, $m_{x_1},m_{x_2}$ the mean value of $x_1$, $x_2$, and $s_{x_1},s_{x_2}$ the standard deviation of  $x_1$, $x_2$. The time lag $\tau$ can be either negative or positive. In general, $C_{x_1,x_2}(\tau)\ne C_{x_2,x_1}(\tau)$, as positive and negative time lags, cover different parts of the waveform signals.

\paragraph{Phase synchronization using the Hilbert transform:}
\label{sec:ht}
The Hilbert transform of a time series $x$ is defined as
\begin{equation}
  H(x)(t)=\frac{1}{\pi}{\rm p.v.}\int_{-\infty}^{\infty}\frac{x(\tau)}{t-\tau}d\tau,
\end{equation}
where ``p.v.'' denotes the Cauchy principal value.
The analytic representation of the time series using the Hilbert transform is:
  $f(t)=x(t)+iH(x)(t)=A(t)e^{i\phi(t)}$.
Given two time series, $x_1(t)$,  $x_2(t)$, with analytic signals $f_{1,2}(t)=A_{1,2}(t)e^{i\phi_{1,2}(t)}$, it is possible to find the relative phase between the time series using 
\begin{equation}
  g(t)=\frac{f_1(t)f_2^*(t)}{A_1(t)A_2(t)}=e^{i(\phi_1(t)-\phi_2(t))}.
\end{equation}
When the two time series are completely synchronized, the phase difference will be constant, and the absolute value of the average (over time) of $g(t)$, $|\langle g(t)\rangle|$, will be one. When the two time series are not synchronized, $|\langle g(t)\rangle|$ will be small and close to zero. Thus, $|\langle g(t)\rangle|$ measures the phase synchronization between time series. The ``envelope'' of the time series is $A(t)$, and its maximal and mean value can provide a measure for the signal’s strength; this is, however, not our main interest in the current study as we focus on the phase difference between the different stations. Still, the magnitude measure can help to identify seismic events. 

\paragraph{The mutual information:}
\label{sec:mi}
We use a symmetric form of mutual information, sometimes called ``symmetric uncertainty''; below, we refer to this relation as mutual information. Given two random variables (or time series), $x_1, x_2$, it is
\begin{equation}
  I_{x_1,x_2}=\frac{2}{H_{x_1}+H_{x_2}}\iint p(x_1,x_2)\log_2\left(\frac{p(x_1,x_2)}{p(x_1)p(x_2)}\right) dx_1dx_2,
\end{equation}
where $p(x_1)$, $p(x_2)$ are the probability density functions (pdfs) of $x_1, x_2$, respectively, $p(x_1,x_2)$ is the joint pdf of $x_1, x_2$, and $H_{x_{1,2}}=\int p(x_{1,2})\log_2p(x_{1,2}) dx_{1,2}$ is the entropy of $x_{1,2}$. When $x_1, x_2$ are independent, the joint probability is $p(x_1,x_2)=p(x_1)p(x_2)$, such that $\log_2\left(\frac{p(x_1,x_2)}{p(x_1)p(x_2)}\right)=0$, and the entire integral is zero. When $p_{x_1}=p_{x_2}$, it is possible to show (using the Dirac $\delta$ function) that the integral of $I$ reduces to the entropy, which eventually yields $I=1$. Thus, the mutual information is bounded between 0 and 1. The mutual information reflects both linear and nonlinear dependencies, contrary to the cross-correlation that reflects only linear dependencies. The mutual information is not affected by the mean and standard deviation of the time series.

\paragraph{The coherence:}
\label{sec:coher}
In contrast to the previous methods, the coherence is calculated in the Fourier space, at each frequency. Given two time series, $x_1$, $x_2$, it is defined as
\begin{equation}
  C_{x_1x_2}(f)=\frac{P_{x_1x_2}}{P_{x_1x_1}P_{x_2x_2}},
\end{equation}
where $P_{x_1x_2}$ is the cross power spectrum density of $x_1$, $x_2$, and $P_{x_1x_1}$ ($P_{x_2x_2}$) is the power spectrum density of $x_1$ ($x_2$). There are two approaches to averaging the spectra. In the first approach, the spectra are calculated using Welch's method \cite[][]{Welch-1967:use} in which the time series is divided into (partially overlapping) segments where each segment is multiplied by a window function, and the spectra of all segments are averaged. In the second approach, the spectra are averaged using a sliding window in the frequency domain. The results shown in the results section were obtained using the first approach. The coherence is maximal when two frequencies match each other, whereas it is zero when there is no overlap. Seismic events are characterized by specific frequencies. The detection of the same frequencies in several stations could indicate the development of an event. We calculated the coherence for each pair of stations and found the maximum coherence within the following frequency ranges: 0.5-1 Hz, 1-5 Hz, 5-10 Hz, 10-20 Hz, and 0.5-20 Hz. The maximum coherence is used to quantify links of the network at different frequency ranges.

The methods described above result in metrics between 0 and 1, where higher values indicate stronger ``connections'' between pairs of stations. These values can serve as network links where the network nodes are the different stations. Below, we describe the ways of characterizing the network.  

\paragraph{$k$-means clustering:}
\label{sec:kmean}
Given a set of values (in our case, a set of link values in a given time window), it is possible to identify different clusters. Given the number of clusters, $k$, the $k$-means clustering method enables the detection of $k$ clusters, such that their similarity within the cluster is maximal, as is the difference (separation/distance) between the different clusters. Within the context of this study, during a seismic event, one expects that the stations that ``felt'' the event would be in the same cluster, while the others would be in different clusters. There are measures that quantify the level of separation between the clusters, in comparison to the similarity within the clusters; one of these measures is the silhouette, and our results (not shown here) indicate that the optimal number of clusters is 2-3.


\subsection{Network construction:}
\label{sec:netwrok-construction} 

We constructed a time-evolving network based on the stations' waveform data using the following steps at each time sample. 
\begin{enumerate}

\item  Each sample spans a $\sim$30 s time window preceding that time sample. For most of the methods (cross-correlation, synchronization, and mutual information), we perform bandpass filtering (typically between 1 and 10 Hz).

\item We apply each of the methods (i.e., cross-correlations, synchronization, mutual information, and coherence) for each pair of stations. We then construct a matrix of connections that represents the link's strength (weights), which is calculated using the different methods described above. 

\item We then compute the global measures of the network links’ matrix (such as the first and second eigenvalues, the degree distribution, and the clustering coefficient); a significant deviation from the background envelope of the measure is assumed to indicate an event.


\item We also perform the $k$-means clustering on the set of links of each method and on the set of vectors that is composed of the links of all methods; when using four methods, the dimension of the vector will be four. Then, the measures, such as the mean link weight within the different clusters and the size of the cluster, are calculated.
  
\end{enumerate}

\paragraph{Detection and verification:}
\label{sec:detec}

The time-dependent links describe a time-evolving network. Each of the global measures of the network results in a time series that reflects the network’s activity level, including the seismic activity in the region. Thus, we expect large deviations from the background to reflect collective behavior or significantly coordinated seismic activity.

We performed the analysis on the waveform data before and after strong earthquakes that occurred within California and in its neighboring states and Mexico because this region has been extensively studied in the past, with a wealth of available seismic station data. Focusing on large events, occurring from 2001{\textendash}2019, helped us to tune the parameters of the proposed procedure and to verify the detection of seismic catalog events against those that were detected by our technique.

\section{Results}
\label{sec:results}

\begin{figure}[t!]
\begin{center}
\includegraphics[width=1.0\textwidth]{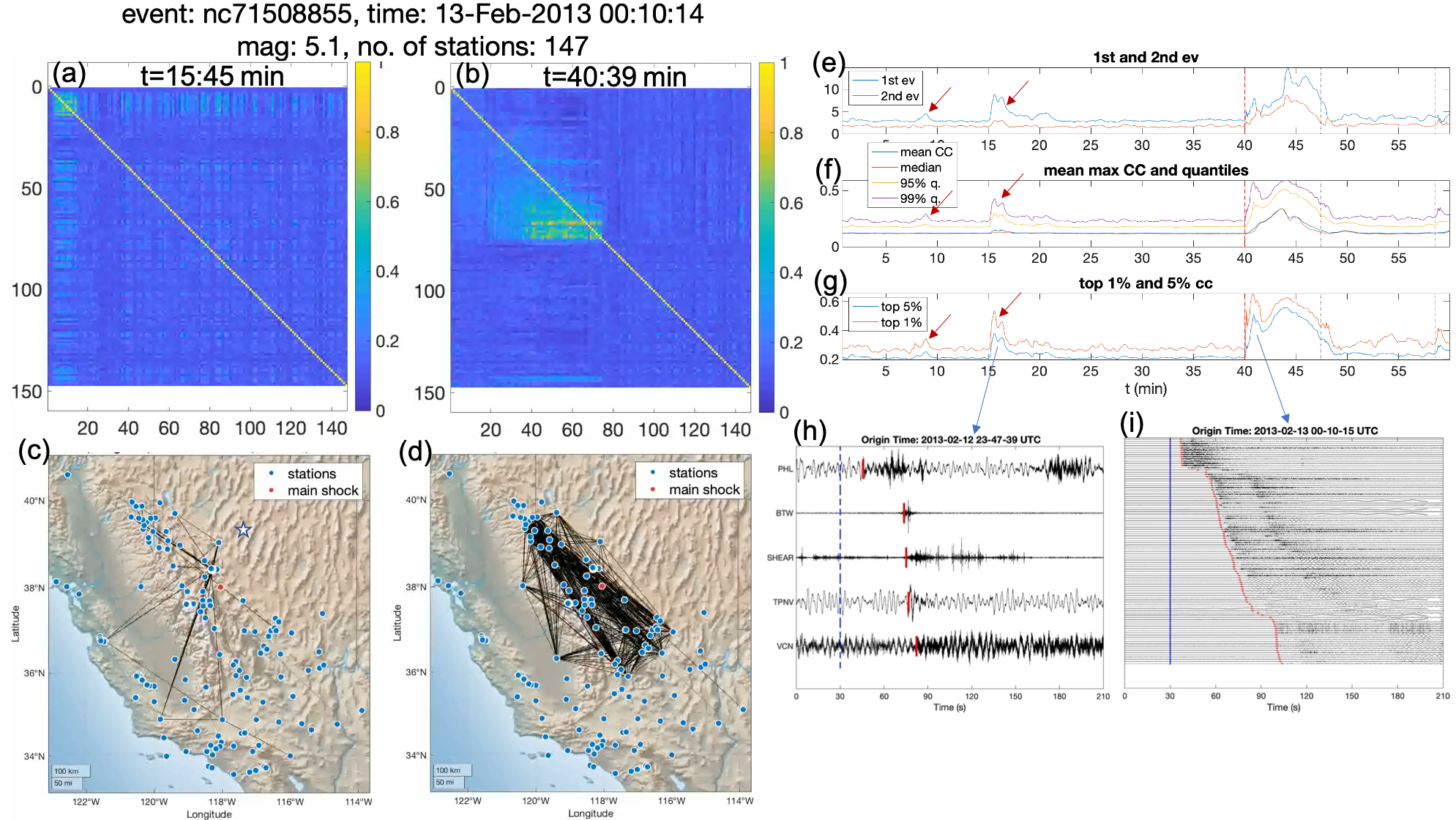}
\end{center}
\caption{Summary of the cross-correlation analysis for the magnitude 5.1 event (indicated by the solid red circle in panels c and d) that occurred in Nevada, close to the California border ($\sim$ 118\degree{W},32\degree{N}) on Feb. 13, 2013, 00:10:14. We analyzed the waveform data of 147 stations (indicated by the solid blue circles in panels c and d).  (a) The matrix of maximum lagged cross-correlations between the different stations for $t$=15:45 min (24:15 minutes prior to the mainshock) at which a small event occurred. The axis numbers indicate the station number where the stations are sorted according to the distance from the mainshock epicenter. (b) Same as (a) for $t$=40:39 min, slightly after the mainshock. Note the region of high cross-correlation values as a result of the earthquake event. (c) Links (black lines) with a maximum cross-correlation larger than 0.4 based on the matrix shown in panel (a). The white star indicates the location of the event that is associated with the increased cross-correlation; the identification and location of the event were based on standard analysis of waveform data. (d) Same as (c) but for the matrix shown in panel (b) during which the mainshock occurred. (e) The absolute value of the first and second eigenvalues of the cross-correlation matrices like the ones shown in panels (a) and (b). The dashed red line indicates the time of the mainshock, while the dotted vertical black lines indicate other events that were included in the California catalog within the time frame of the figure. The red arrows indicate increased measures that are linked to events that were found by analyzing the waveform data and that were not included in the California catalog. (f) Same as panel (e) but for the mean of the cross-correlation matrix (excluding the diagonal), median, 95\% quantile, and 99\% quantile. (g) Same as (e) for the mean top 1\% and 5\% of the cross-correlation matrix. (h) The waveforms of the unreported event that is indicated by the blue arrow in panel (g). (i) The waveforms of the mainshock, indicated by the blue arrow in panel (g). }\label{fig:detection}
\end{figure}

Below, we provide an example of the use of the network approach (here using the lagged cross-correlation) to detect real-time events that were not included in the catalog. We focused on the 5.1 magnitude event that occurred on Feb. 13, 2013, 00:10:14, in Nevada near the California border ($\sim$ 118\degree{W}, 32\degree{N}). We calculated the lagged cross-correlation function of 147 stations (nodes) around the event epicenter. We considered the 40 minutes preceding the event and 20 minutes after the event, where the cross-correlation function for each pair of links was calculated every 6 seconds based on the preceding 30 seconds. Then, we constructed a matrix of the maximum lagged cross-correlation values and analyzed this matrix. Examples of such matrices are shown in Fig. \ref{fig:detection}a,b. These matrices are associated with the mainshock (Fig. \ref{fig:detection}b) and a foreshock (Fig. \ref{fig:detection}a) that was not included in the catalog. The links with a maximum cross-correlation larger than 0.4 are indicated in Fig. \ref{fig:detection}c,d (black lines) and represent the extent of the events at the given time. Changing the cross-correlation threshold to a value other than 0.4 would result in a different link distribution, but it is clear that the number of links increases during earthquake events. In Fig. \ref{fig:detection}e-g, we plot several global measures of the cross-correlation matrices (like the ones of Fig. \ref{fig:detection}a,b). These include the first and second eigenvalues of the matrices (Fig. \ref{fig:detection}e), the mean value and several quantiles of the maximum cross-correlation matrices (Fig. \ref{fig:detection}f), and the mean of the top 5\% and 1\% of the matrix values (Fig. \ref{fig:detection}f). The mainshock is indicated by the dashed vertical red line, while other events of the catalog are indicated by the dotted vertical black lines. We identified two events that were not included in the catalog; these are indicated by the red arrows. The waveform data (Fig. \ref{fig:detection}h) verified that this is indeed a real event. We also located the event and found that it is within the study area (white star in Fig. \ref{fig:detection}c).

\begin{figure}[t!]
\begin{center}
\includegraphics[width=0.7\textwidth]{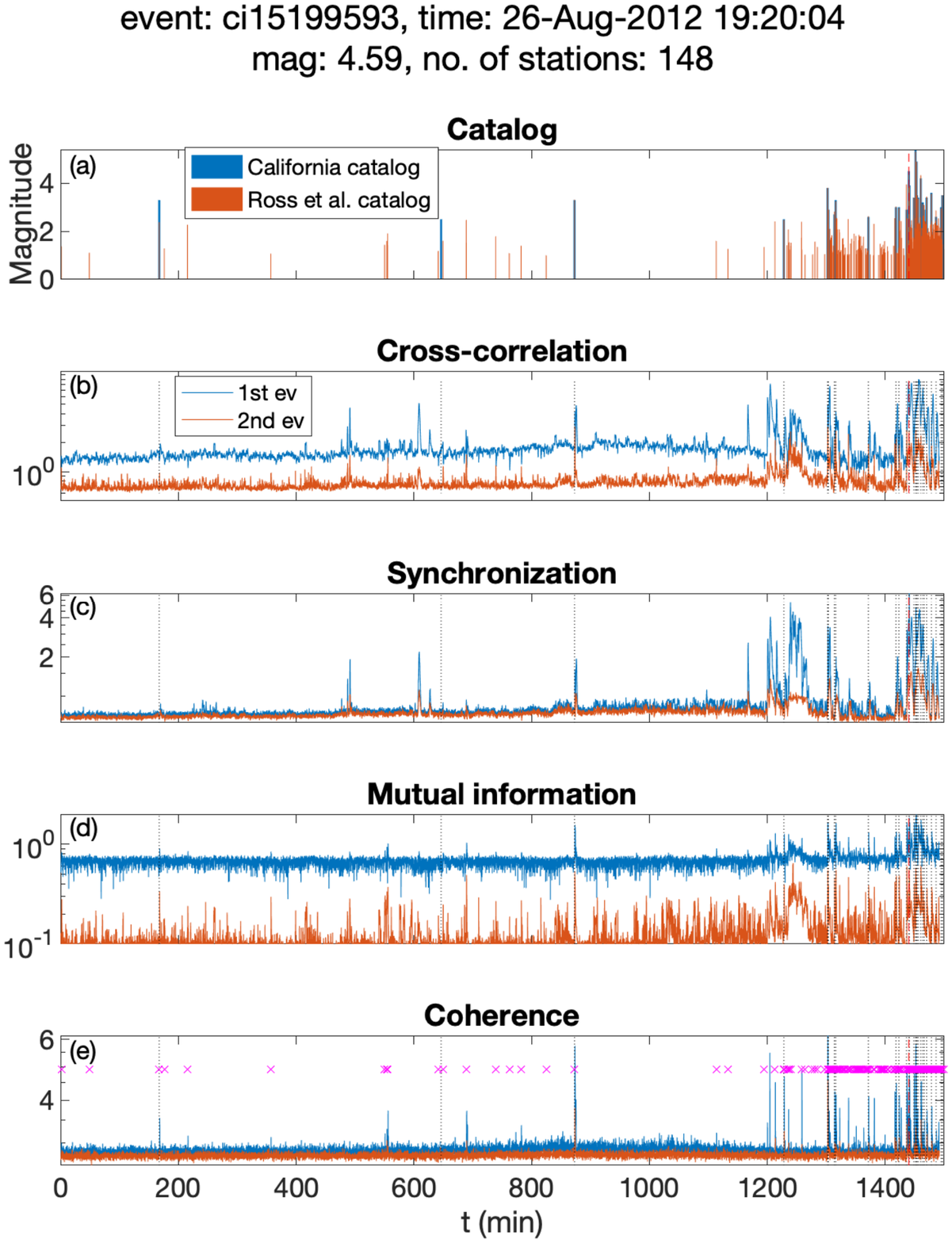}
\end{center}
\caption{A 25-hour analysis of the different methods. We consider a 4.59 $M_w$ event that occurred on August 26, 2012, 19:20:04 at 33\degree{N}, 115.5\degree{W}. We analyzed 148 stations’ data starting one day before the event to one hour after the event. (a) Events included in the California catalog (blue bars) and the Ross {\it et al.} \cite{Ross-Trugman-Hauksson-et-al-2019:searching} catalog (red bars) versus their corresponding magnitudes. The mainshock is indicated by the dashed vertical red line. The $x$ axis indicates the time (in minutes) since one day before the mainshock, i.e., since August 25, 2012, 19:20:04. The first and second eigenvalues versus time for the different methods are depicted in panels b to e as follows: (b) maximum lagged cross-correlation matrix, (c) synchronization, (d) mutual information, and (e) coherence. The dotted vertical black lines in panels b{\textendash}e indicate events that are included in the California catalog, and the pink ``$\times$''s in panel e indicate events that are included in the Ross {\it et al.} \cite{Ross-Trugman-Hauksson-et-al-2019:searching} catalog. Note the match between the events in the catalog and the different measures’ peaks, as well as the match between the peaks in the different methods’ curves (and especially those of the coherence method) and the events included in the Ross {\it et al.} \cite{Ross-Trugman-Hauksson-et-al-2019:searching} catalog. Note also that not all the events included in the Ross {\it et al.} \cite{Ross-Trugman-Hauksson-et-al-2019:searching} catalog coincide with peaks of the coherence curve as some of these events are outside the region covered by the spread of the analyzed stations. }\label{fig:methods}
\end{figure}

We summarize the results of the different methods in Fig. \ref{fig:methods}. Here we focus on a 4.59 $M_w$ earthquake that occurred on August 26, 2012, at 19:20:04, 4 km north of Brawley, CA (33\degree{N}, 115.6\degree{W}). This event was a foreshock for a larger 5.4 $M_w$ event that occurred in the same region 11 minutes later. We considered data from 148 stations within a radius of 5 degrees from the epicenter of the event. We analyzed the data 24 hours before the event and one hour after. We first present the California catalog (which is complete above magnitude 2.5) and show that there were several events one day before the main event (Fig. \ref{fig:methods}a). We also present events included in the catalog of Ross {\it et al.} \cite{Ross-Trugman-Hauksson-et-al-2019:searching}, which is complete above magnitude 1. As expected, there were many smaller events within one day before the mainshock. In both catalogs, the rate of the events increased several hours before the main event. In Fig. \ref{fig:methods}b-e, we present the first and second eigenvalues of the link matrices (like the ones shown in Fig. \ref{fig:detection}a,b) as a function of time. These were calculated at each time using the cross-correlation, synchronization, mutual information, and coherence methods. In these panels, the main event is marked by the dashed vertical red line while the other events of the California catalog are presented by the dashed vertical black lines. Almost all these events coincide with peaks in the graphs of the different measures, thus suggesting the detection power of the different techniques. Yet, there are additional peaks that do match the events in the main catalog---these seem to match the smaller magnitude events that are included in the catalog of Ross {\it et al.} \cite{Ross-Trugman-Hauksson-et-al-2019:searching}; see the coherence results in Fig. \ref{fig:methods}e. [Note that the epicenters of the small events that do not match the peaks in the curves may be outside the area of the stations we analyzed.] Thus, the network approach suggested here has the potential to reveal, in real-time, small-magnitude events.

\begin{figure}[t!]
\begin{center}
\includegraphics[width=0.8\textwidth]{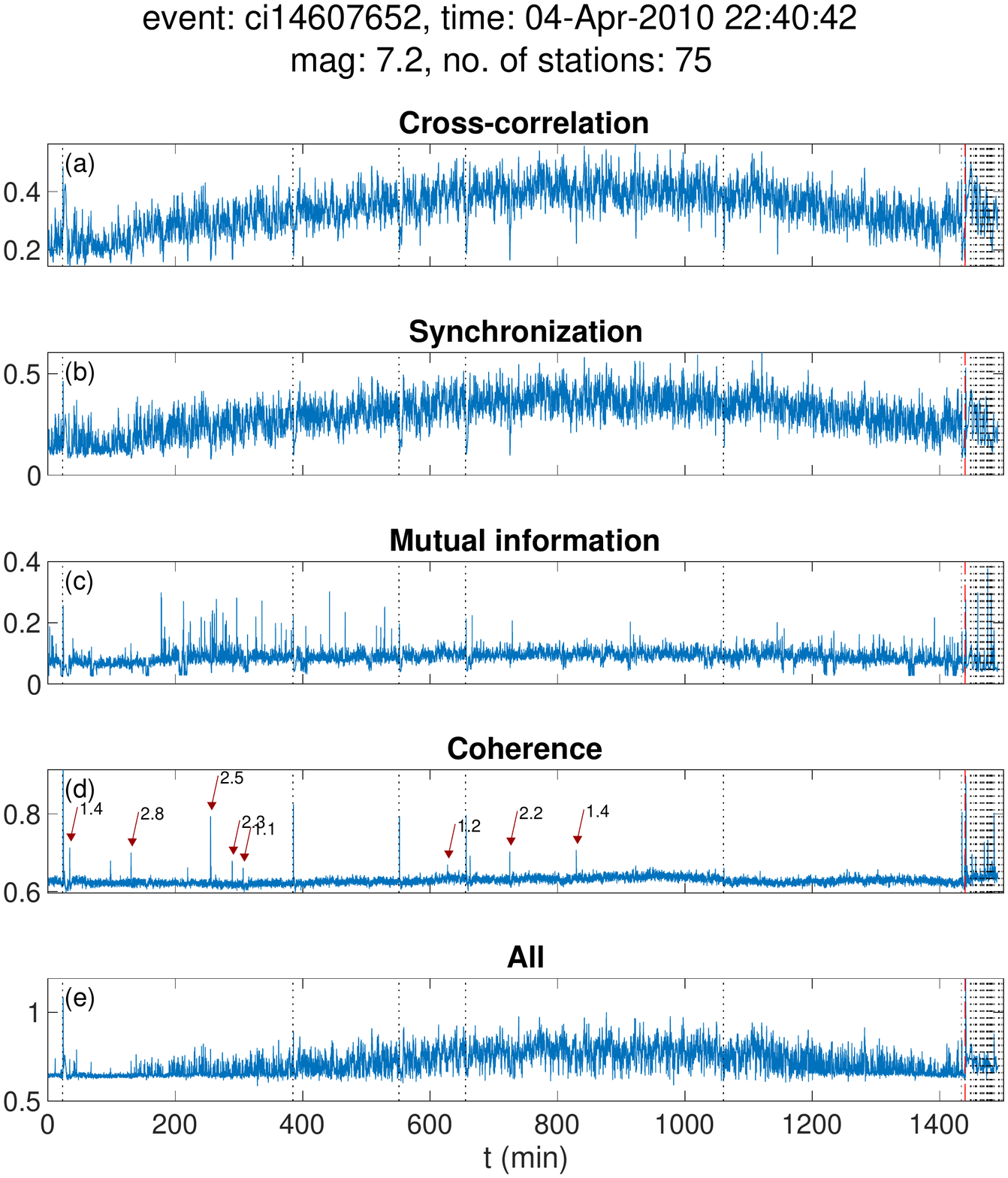}
\end{center}
\caption{A 25-hour analysis of the $k$-means clustering of the different methods. We consider a 7.2 $M_w$ event that occurred on April 4, 2010, 22:40:42 at 32.3\degree{N}, 115.3\degree{W}, in Baja California, Mexico. We analyzed 75 stations’ (see Fig. \ref{fig:clusters-map}b) data starting one day before the event to one hour after the event. For the clustering analysis, we used two clusters that are based on the links (``connections'') quantified by the different methods, and we computed the mean value within the most dominant cluster. The figure depicts this mean calculated based on the following methods: (a) cross-correlation, (b) synchronization, (c) mutual information, and (d) coherence. (e) The mean vector length within the dominant cluster, which is based on the analysis of a four-dimensional vector composed of all methods. The dashed vertical red line indicates the mainshock, while the dotted vertical lines indicate the other events included in the California catalog. Note the match between seismic events and the peaks (or ``pits'') and the times of the events. In panel (d), we indicate with arrows the peaks in the coherence measure that may be associated with events from the Ross {\it et al.} \cite{Ross-Trugman-Hauksson-et-al-2019:searching} catalog. }\label{fig:clusters}
\end{figure}

The graphs of the different methods exhibit peaks that are not associated with events in either catalog. Moreover, there are periods of significant enhancement [e.g., in the range of $t=1200-1300$ minutes--this is not present in the curves of the coherence method (Fig. \ref{fig:methods}e)]. Such a pattern is most probably not related to seismic activity. In addition, a diurnal cycle is visible in the first eigenvalue curve of the cross-correlation (Fig. \ref{fig:methods}e) and, to a lesser extent, the coherence curve (Fig. \ref{fig:methods}b). Since seismic events have a short time scale, it is easy to separate the peaks that may be associated with seismic events from the background envelope. Generally speaking, the coherence method exhibits the best results. 

\begin{figure}[t!]
\begin{center}
\includegraphics[width=0.8\textwidth]{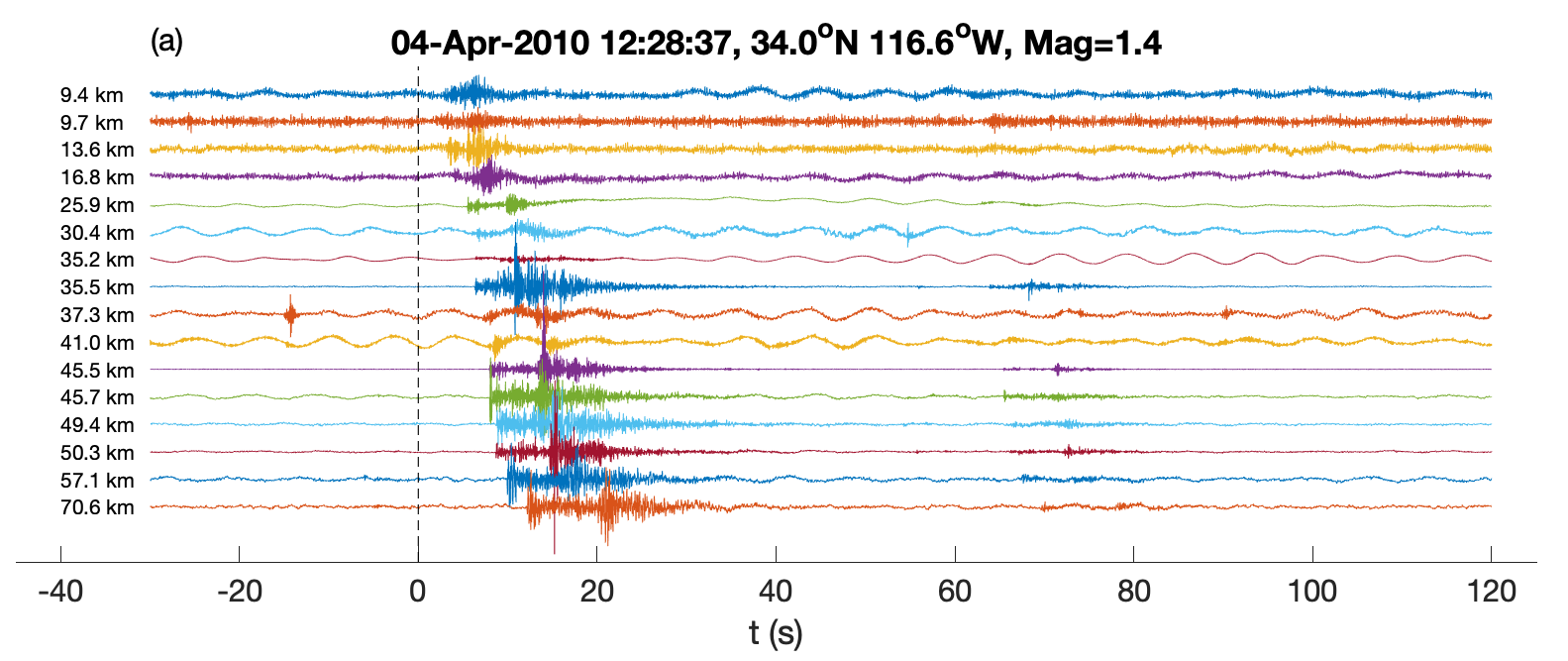}\\
\includegraphics[width=0.49\textwidth]{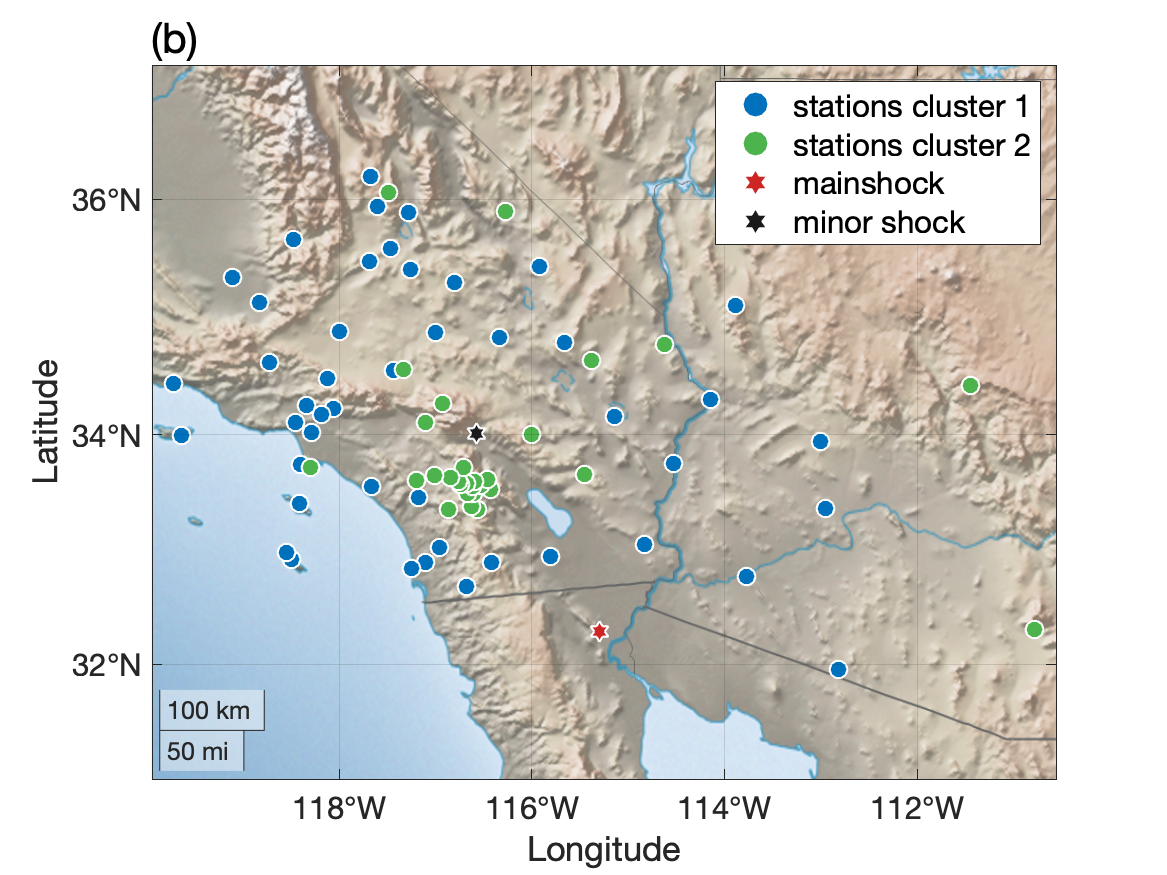}
\includegraphics[width=0.5\textwidth]{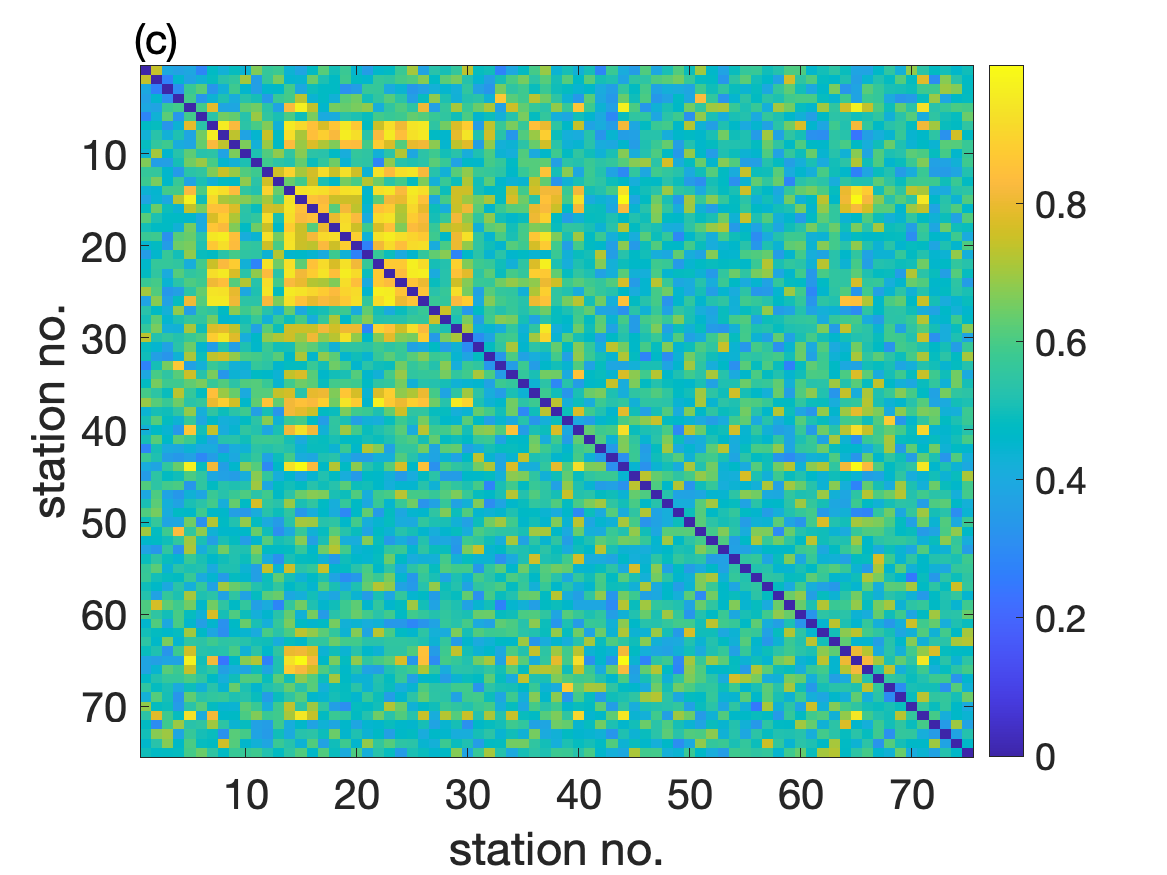}
\end{center}
\caption{Analysis of a minor 1.4 $M_w$ event that occurred on April 4, 2010, 12:28:37 at 34\degree{N} 116.6\degree{W}, before the mainshock presented in Fig. \ref{fig:clusters}; this event is indicated by the rightmost red arrow in Fig. \ref{fig:clusters}d. (a) The waveforms of stations close to the epicenter of the event--the distance of the stations from the epicenter is indicated on the left side. The dashed vertical black line indicates the time of the event. (b) The distribution of the stations from the analysis in Fig. \ref{fig:clusters} (full circles). The green circles indicate the nodes that have more than 24 links from the dominant cluster of the coherence method--see Fig. \ref{fig:clusters}d. The red and black hexagrams indicate the epicenters of the mainshock and the minor shock. (c) The matrix of links that are based on the coherence method. Note the cluster of high coherence on the upper left part. The stations are ordered by their distance from the mainshock’s epicenter.  }\label{fig:clusters-map}
\end{figure}

An additional earthquake sequence, with $M_w$ of 7.2, was examined and is presented here. This event occurred on April 4, 2010, 22:40:42, 12 km southwest of Delta, Baja California, Mexico (32.3\degree{N}, 115.3\degree{W}). Here, the analysis was based on data from 75 stations within a radius of 5 degrees from the epicenter of the mainshock; all stations are USA stations, and all of them (except one) are located north of the epicenter (Fig. \ref{fig:clusters-map}b). Similar to Fig. \ref{fig:methods}, we studied the 24 hours before the mainshock and one hour after the mainshock. Here, we performed the $k$-means clustering analysis with two clusters using the four suggested methods, namely, cross-correlation, synchronization, mutual information, and coherence. In addition, we performed the cluster analysis with the four-dimensional vector whose components are the values of the four methods. We normalized each of the methods' outputs to one, to allow unbiased analysis. At each time sample point, we present the mean value of the cluster with the larger values--see Fig. \ref{fig:clusters}. The events that are included in the California catalog are marked by the dashed vertical lines and coincide nicely with the peaks (or ``pits'') of the different curves. The diurnal cycle is present in the cross-correlation and synchronization curves, and to a lesser degree, in the mutual information curve. Yet, the peaks (or pits) are detectable. The additional peaks that do not match the California catalog events may represent smaller magnitude events. In Fig. \ref{fig:clusters}d, we present the coherence results, and the red arrows indicate the peaks that may be associated with events from the Ross {\it et al.} \cite{Ross-Trugman-Hauksson-et-al-2019:searching} catalog; the event magnitudes are listed next to the arrows. We focused on one of these events, the rightmost one with a magnitude of 1.4; the waveforms of this event are shown in Fig. \ref{fig:clusters-map}a. In Fig. \ref{fig:clusters-map}c, we show the matrix that represents the links’ strength using the coherence method. A clear cluster is visible in this plot. In Fig. \ref{fig:clusters-map}b, stations with degree nodes (i.e., the number of links associated with the node) larger than 24 are marked by green circles. The epicenter of this 1.4 event is indicated by the black hexagram, and it is surrounded by the stations (nodes) of the main cluster. There are ``green'' stations (nodes) far from the epicenter of the minor event, but their number is small. The clustering analysis, based on all methods (Fig. \ref{fig:clusters}e), combines these four methods’ features, and its performance is rather poor. It is clear, following Fig. \ref{fig:clusters}, that the coherence results exhibit the best performance. 

Similar to the above, we analyzed 27 time series (each spans a time window of 25 hours, 24 before a mainshock and one hour after) using the k-mean clustering with three clusters using the coherence analysis; these include the events shown and discussed above. Fig. \ref{fig:map_all_events} shows the locations of these mainshocks while Figs. \ref{fig:event_stations_map1}--\ref{fig:event_stations_map3} show the locations of stations that were used in the analysis for each of the time series (that are associated with different mainshocks). We expect an increase in the coherence when an earthquake above a certain magnitude occurs in the vicinity of the analyzed stations. For this reason, we also mark in Figs. \ref{fig:event_stations_map1}--\ref{fig:event_stations_map3} the timing of the earthquakes that are included in the California catalog (marked by black $\times$s) and in the Ross {\it et al.} \cite{Ross-Trugman-Hauksson-et-al-2019:searching} catalog (marked by pink $\times$s). We mark the events that occur within the black ellipses drawn in Figs. \ref{fig:event_stations_map1}--\ref{fig:event_stations_map3}. The ellipse equation is $(x-\bar{x})^2/\sigma_x^2+(y-\bar{y})^2/\sigma_y^2=1.5^2$ where $\bar{x}$ and $\bar{y}$ are the mean zonal and meridional station locations and $\sigma_x^2$ and $\sigma_y^2$ are the zonal and meridional variances of the station locations. This procedure aims at excluding events that occur far away from the majority of the stations.

As mentioned above, we expect an increase in the coherence measure when an earthquake event occurs in the vicinity of the stations. To verify this, we compare events that are included in the California catalog and in the Ross {\it et al.} \cite{Ross-Trugman-Hauksson-et-al-2019:searching} catalog to peaks in the coherence time series. Peaks that match the earthquake events validate the hypothesis that increased coherence is associated with enhanced seismic activity. However, earthquakes that do not match peaks in the coherence time series invalidate the hypothesis. Yet, peaks in the coherence time series that do not match earthquakes that are included in the catalogs can represent events that are outside the regional extent of the catalogs and thus do not invalidate the hypothesis. In some of the cases (e.g., Fig.~\ref{fig:cluster3}g), there is an excellent match between earthquakes and coherence peaks; in the case of Fig.~\ref{fig:cluster3}g, the marked earthquakes (indicated by the dashed vertical lines) have magnitudes larger than 2.5. In fact, the vast majority of the these California catalog events (whose magnitudes are larger than 2.5) match coherence peaks (Figs. \ref{fig:cluster1}-\ref{fig:cluster3} that contain the dashed vertical lines). Thus, we can state that the coherence measure captures earthquake events with magnitudes larger than 2.5. As for the weaker events (magnitudes larger than 1.3) of the Ross {\it et al.} catalog \cite{Ross-Trugman-Hauksson-et-al-2019:searching}, there is, in some cases, a good match between the peaks in the coherence times series and the events. For example, almost every event indicated in Fig. \ref{fig:cluster1}a can be matched with a peak in the coherence; some of these peaks are small but noticeable. In other cases, like in Fig. \ref{fig:cluster1}g, there are several events (of the Ross {\it et al.} \cite{Ross-Trugman-Hauksson-et-al-2019:searching} catalog) that do not match peaks in the coherence time series; these might be the ones that are simply adjusted to the border of the ellipse (see Fig.~\ref{fig:event_stations_map1}g). There are more cases for which events of the Ross {\it et al.} \cite{Ross-Trugman-Hauksson-et-al-2019:searching} catalog are not associated with peaks in the coherence time series. We thus conclude that the coherence measure reflects quite reasonably earthquake events with magnitudes larger than 2.5 but that it is less successful in capturing the weaker events included in the Ross {\it et al.} \cite{Ross-Trugman-Hauksson-et-al-2019:searching} catalog. 

\section{Summary and Discussion.}\label{sec4}

We used a network approach to study earthquake activity. More specifically, we studied stations' waveform data 24 hours before mainshocks and one hour after. At each time point, we constructed a network whose nodes are the stations and whose links are quantified by different statistical measures, including cross-correlation, synchronization, mutual information, and coherence. We show that the collective network link measures are significantly different (often significantly larger) during earthquakes. We also applied the $k$-means clustering method on the different measures and focused on the properties of the largest cluster---this procedure aims at capturing the development of the evolving seismic activity. We also constructed the cluster based on a vector composed of all measures. We found that the coherence measure exhibits the best correspondence with earthquake events, where earthquake events are matched with peaks of the coherence. The performance is good for earthquakes whose magnitudes are larger than 2.5 and moderate for earthquakes with magnitudes larger than 1.3.

While we have demonstrated that earthquake activity may be captured using collective network measures of seismic stations’ waveforms, it is not clear that these measures can improve the techniques that are currently used for earthquake detection and early warnings. Yet, it is possible that a more careful analysis of the network may improve the detection performance. One possible idea is to cover the study region with overlapping circles/ellipses and to construct an evolving network of each sub-region. Each network may reflect the activity of the sub-region, and strong (different) enough measures may reflect seismic activity within the more localized sub-regions. In addition, it is possible that other collective network measures may better capture earthquake activity. 

\section*{Acknowledgments}
We thank the European Research Executive Agency project no. 821115 (RISE) and the Israel Ministry of Energy (grant no. 221-17-017) for financial support. We thank Golan Bel for the helpful discussions. 

\begin{table}[!ht]
    \centering
    \caption{Information on the 27 events considered in this paper.}
    \label{table:events}
    \begin{tabular}{c|c|c|c|c|c|c}
    \hline
        i.d. & Time & Latitude & Longitude & \makecell{Depth \\(km)} & Magnitude & \makecell{No.\\stations} \newline
    \\ \hline \hline
        ci11066877 & 2012-02-14 10:09:28 & 32.2 & -115.21 & 9.963 & 4.87 mw & 131 \\ \hline
        ci11129826 & 2012-07-01 03:25:20 & 32.21 & -115.34 & 2.821 & 4.64 mw & 90 \\ \hline
        ci14607652 & 2010-04-04 22:40:42 & 32.29 & -115.29 & 9.987 & 7.2 mw & 75 \\ \hline
        ci14937372 & 2011-02-18 17:47:35 & 32.05 & -115.06 & 14.929 & 5.09 mw & 127 \\ \hline
        ci15199593 & 2012-08-26 19:20:04 & 33.02 & -115.54 & 4.79 & 4.59 mw & 148 \\ \hline
        ci15296281 & 2013-03-11 16:56:06 & 33.50 & -116.46 & 10.943 & 4.7 mh & 137 \\ \hline
        ci37374687 & 2016-06-10 08:04:38 & 33.43 & -116.44 & 12.31 & 5.19 mw & 98 \\ \hline
        ci38138175 & 2018-11-19 20:18:42 & 32.21 & -115.25 & 10.95 & 4.82 mw & 87 \\ \hline
        ci38443183 & 2019-07-04 17:33:49 & 35.71 & -117.5 & 10.5 & 6.4 mw & 123 \\ \hline
        nc71508850 & 2011-01-12 08:51:03 & 36.77 & -121.5 & 7.661 & 4.5 mw & 104 \\ \hline
        nc71508851 & 2011-04-17 00:45:37 & 38.39 & -118.74 & 5.1 & 4.6 ml & 137 \\ \hline
        nc71508852 & 2011-08-27 07:18:21 & 36.58 & -121.18 & 7.5 & 4.64 mw & 116 \\ \hline
        nc71508854 & 2012-10-21 06:55:09 & 36.31 & -120.86 & 8.619 & 5.29 mw & 152 \\ \hline
        nc71508855 & 2013-02-13 00:10:14 & 38.02 & -118.06 & 7.5 & 5.1 ml & 141 \\ \hline
        nc71508856 & 2013-05-29 14:38:03 & 34.41 & -119.92 & 7.12 & 4.8 mw & 128 \\ \hline
        nc71508857 & 2014-03-29 04:09:42 & 33.9325 & -117.92 & 5.09 & 5.1 mw & 152 \\ \hline
        nc71508858 & 2014-07-05 16:59:34 & 34.28 & -117.03 & 7.25 & 4.58 mw & 152 \\ \hline
        nc71508859 & 2014-08-24 10:20:44 & 38.22 & -122.31 & 11.12 & 6.02 mw & 152 \\ \hline
        nc71508860 & 2015-02-14 02:28:56 & 37.14 & -117.26 & 17 & 4.8 ml & 126 \\ \hline
        nc71508861 & 2016-02-16 23:04:26 & 37.20 & -118.4 & 15.06 & 4.77 mw & 116 \\ \hline
        nc71508863 & 2016-07-07 23:40:45 & 38.65 & -118.79 & 9.1 & 4.5 ml & 129 \\ \hline
        nc71508864 & 2016-12-28 08:18:00 & 38.38 & -118.9 & 11.3 & 5.6 ml & 42 \\ \hline
        nc71508867 & 2017-11-13 19:31:29 & 36.63 & -121.24 & 6.31 & 4.58 mw & 100 \\ \hline
        nc71508868 & 2018-07-05 15:17:36 & 35.89 & -115.78 & 8.6 & 4.5 ml & 117 \\ \hline
        nc71508875 & 2019-07-06 03:19:53 & 35.77 & -117.6 & 8 & 7.1 mw & 75 \\ \hline
        nc71508904 & 2019-10-15 19:42:30 & 36.65 & -121.27 & 10.1 & 4.71 mw & 101 \\ \hline
        nc71508905 & 2020-01-25 03:03:34 & 35.1 & -116.97 & 3.07 & 4.62 mw & 115 \\ \hline
    \end{tabular}
\end{table}





\bibliographystyle{unsrt}

\bibliography{all}

\begin{thebibliography}{10}

\bibitem{Freddi-Galasso-Cremen-et-al-2021:innovations}
Fabio Freddi, Carmine Galasso, Gemma Cremen, Andrea Dall’Asta, Luigi
  Di~Sarno, Agathoklis Giaralis, Fernando Guti{\'e}rrez-Urz{\'u}a, Christian
  M{\'a}laga-Chuquitaype, Stergios~A Mitoulis, Crescenzo Petrone, et~al.
\newblock Innovations in earthquake risk reduction for resilience: Recent
  advances and challenges.
\newblock {\em International Journal of Disaster Risk Reduction}, 60:102267,
  2021.

\bibitem{Kuyuk-Allen-2013:optimal}
H~Serdar Kuyuk and Richard~M Allen.
\newblock Optimal seismic network density for earthquake early warning: A case
  study from california.
\newblock {\em Seismological Research Letters}, 84(6):946--954, 2013.

\bibitem{Gasparini-Manfredi-Zschau-2007:earthquake}
Paolo Gasparini, Gaetano Manfredi, and Jochen Zschau, editors.
\newblock {\em Earthquake early warning systems}.
\newblock Springer, 2007.

\bibitem{Jordan-Chen-Gasparini-et-al-2011:operational}
Thomas~H Jordan, Yun-Tai Chen, Paolo Gasparini, Raul Madariaga, Ian Main,
  Warner Marzocchi, Gerassimos Papadopoulos, Gennady Sobolev, Koshun Yamaoka,
  and Jochen Zschau.
\newblock Operational earthquake forecasting. state of knowledge and guidelines
  for utilization.
\newblock {\em Annals of Geophysics}, 54(4), 2011.

\bibitem{Ogata-1988:statistical}
Yosihiko Ogata.
\newblock Statistical models for earthquake occurrences and residual analysis
  for point processes.
\newblock {\em Journal of the American Statistical Association}, 83(401):9--27,
  1988.

\bibitem{Gutenberg-Richter-1944:frequency}
Beno Gutenberg and Charles~F Richter.
\newblock Frequency of earthquakes in california.
\newblock {\em Bull. Seismol. Soc. Am.}, 34(4):185--188, 1944.

\bibitem{Gutenberg-Richter-1956:earthquake}
Beno Gutenberg and Carl~F Richter.
\newblock Earthquake magnitude, intensity, energy, and acceleration: (second
  paper).
\newblock {\em Bull. Seismol. Soc. Am.}, 46(2):105--145, 1956.

\bibitem{Omori-1894:after}
Fusakichi Omori.
\newblock {\em On the aftershocks of earthquakes}, volume~7.
\newblock Imperial University of Tokyo, 1894.

\bibitem{Utsu-1961:statistical}
Tokuji Utsu.
\newblock A statistical study on the occurrence of aftershocks.
\newblock {\em Geophys. Mag.}, 30:521--605, 1961.

\bibitem{Ogata-1998:space}
Yosihiko Ogata.
\newblock Space-time point-process models for earthquake occurrences.
\newblock {\em Annals of the Institute of Statistical Mathematics},
  50(2):379--402, 1998.

\bibitem{Gerstenberger-Wiemer-Jones-et-al-2005:real}
Matthew~C Gerstenberger, Stefan Wiemer, Lucile~M Jones, and Paul~A Reasenberg.
\newblock Real-time forecasts of tomorrow's earthquakes in {California}.
\newblock {\em Nature}, 435(7040):328--331, 2005.

\bibitem{Taroni-Marzocchi-Schorlemmer-et-al-2018:prospective}
Matteo Taroni, Warner Marzocchi, Danijel Schorlemmer, Maximilian~Jonas Werner,
  Stefan Wiemer, Jeremy~Douglas Zechar, Lukas Heiniger, and Fabian Euchner.
\newblock Prospective {CSEP} evaluation of 1-day, 3-month, and 5-yr earthquake
  forecasts for {Italy}.
\newblock {\em Seismol. Res. Lett.}, 89(4):1251--1261, 2018.

\bibitem{Woessner-Hainzl-Marzocchi-et-al-2011:retrospective}
J~Woessner, Sebastian Hainzl, W~Marzocchi, MJ~Werner, AM~Lombardi, F~Catalli,
  B~Enescu, M~Cocco, MC~Gerstenberger, and S~Wiemer.
\newblock A retrospective comparative forecast test on the 1992 {Landers}
  sequence.
\newblock {\em J. Geophys. Res.: Solid Earth}, 116(B5), 2011.

\bibitem{Ross-Trugman-Hauksson-et-al-2019:searching}
Zachary~E Ross, Daniel~T Trugman, Egill Hauksson, and Peter~M Shearer.
\newblock Searching for hidden earthquakes in {Southern} {California}.
\newblock {\em Science}, 364(6442):767--771, 2019.

\bibitem{Tan-Waldhauser-Ellsworth-et-al-2021:machine}
Yen~Joe Tan, Felix Waldhauser, William~L Ellsworth, Miao Zhang, Weiqiang Zhu,
  Maddalena Michele, Lauro Chiaraluce, Gregory~C Beroza, and Margarita Segou.
\newblock Machine-learning-based high-resolution earthquake catalog reveals how
  complex fault structures were activated during the 2016--2017 central italy
  sequence.
\newblock {\em The Seismic Record}, 1(1):11--19, 2021.

\bibitem{Cohen-Havlin-2010:complex}
Reuven Cohen and Shlomo Havlin.
\newblock {\em Complex networks: structure, robustness and function}.
\newblock Cambridge university press, 2010.

\bibitem{Tsonis-Roebber-2004:architecture}
Anastasios~A Tsonis and Paul~J Roebber.
\newblock The architecture of the climate network.
\newblock {\em Physica A}, 333:497--504, 2004.

\bibitem{Zhang-Fan-Chen-et-al-2019:significant}
Yongwen Zhang, Jingfang Fan, Xiaosong Chen, Yosef Ashkenazy, and Shlomo Havlin.
\newblock Significant impact of {Rossby} waves on air pollution detected by
  network analysis.
\newblock {\em Geophys. Res. Lett.}, 46(21):12476--12485, 2019.

\bibitem{Zhang-Hu-Kurths-et-al-2013:explosive}
Xiyun Zhang, Xin Hu, J~Kurths, and Zonghua Liu.
\newblock Explosive synchronization in a general complex network.
\newblock {\em Phys. Rev. E}, 88(1):010802, 2013.

\bibitem{Schaff-Waldhauser-2010:one}
David~P Schaff and Felix Waldhauser.
\newblock One magnitude unit reduction in detection threshold by cross
  correlation applied to {Parkfield} ({California}) and {China} seismicity.
\newblock {\em Bull. Seismol. Soc. Am.}, 100(6):3224--3238, 2010.

\bibitem{Gao-Kao-2020:optimization}
Dawei Gao and Honn Kao.
\newblock Optimization of the match-filtering method for robust repeating
  earthquake detection: The multisegment cross-correlation approach.
\newblock {\em J. Geophys. Res.: Solid Earth}, 125(7):e2020JB019714, 2020.

\bibitem{Hauksson-Shearer-2005:southern}
Egill Hauksson and Peter Shearer.
\newblock Southern {California} hypocenter relocation with waveform
  cross-correlation, part 1: {Results} using the double-difference method.
\newblock {\em Bull. Seismol. Soc. Am.}, 95(3):896--903, 2005.

\bibitem{Shearer-Hauksson-Lin-2005:southern}
Peter Shearer, Egill Hauksson, and Guoqing Lin.
\newblock Southern {California} hypocenter relocation with waveform
  cross-correlation, {Part} 2: {Results} using source-specific station terms
  and cluster analysis.
\newblock {\em Bull. Seismol. Soc. Am.}, 95(3):904--915, 2005.

\bibitem{Snieder-Miyazawa-Slob-et-al-2009:comparison}
Roel Snieder, Masatoshi Miyazawa, Evert Slob, Ivan Vasconcelos, and Kees
  Wapenaar.
\newblock A comparison of strategies for seismic interferometry.
\newblock {\em Surv. Geophys.}, 30(4-5):503--523, 2009.

\bibitem{Wapenaar-Van-Ruigrok-et-al-2011:seismic}
Kees Wapenaar, Joost Van Der~Neut, Elmer Ruigrok, Deyan Draganov, J{\"u}rg
  Hunziker, Evert Slob, Jan Thorbecke, and Roel Snieder.
\newblock Seismic interferometry by crosscorrelation and by multidimensional
  deconvolution: A systematic comparison.
\newblock {\em Geophysical J. Inter.}, 185(3):1335--1364, 2011.

\bibitem{Roux-Sabra-Gerstoft-et-al-2005:p}
Philippe Roux, Karim~G Sabra, Peter Gerstoft, WA~Kuperman, and Michael~C
  Fehler.
\newblock P-waves from cross-correlation of seismic noise.
\newblock {\em Geophys. Res. Lett.}, 32:L19303, 2005.

\bibitem{Shapiro-Campillo-Stehly-et-al-2005:high}
Nikolai~M Shapiro, Michel Campillo, Laurent Stehly, and Michael~H Ritzwoller.
\newblock High-resolution surface-wave tomography from ambient seismic noise.
\newblock {\em Science}, 307(5715):1615--1618, 2005.

\bibitem{Cansi-Plantet-Massinon-1993:earthquake}
Yves Cansi, Jean-Louis Plantet, and Bernard Massinon.
\newblock Earthquake location applied to a mini-array: k-spectrum versus
  correlation method.
\newblock {\em Geophys. Res. Lett.}, 20(17):1819--1822, 1993.

\bibitem{Eisermann-Ziv-Wust-Bloch-2018:array}
Andreas~S Eisermann, Alon Ziv, and Hillel~G Wust-Bloch.
\newblock Array-based earthquake location for regional earthquake early
  warning: Case studies from the dead sea transformarray-based earthquake
  location for regional earthquake early warning.
\newblock {\em Bull. Seismol. Soc. Am.}, 108(4):2046--2053, 2018.

\bibitem{Ruigrok-Gibbons-Wapenaar-2017:cross}
Elmer Ruigrok, Steven Gibbons, and Kees Wapenaar.
\newblock Cross-correlation beamforming.
\newblock {\em J Seismol.}, 21(3):495--508, 2017.

\bibitem{Shi-Seydoux-Poli-2021:unsupervised}
Peidong Shi, L{\'e}onard Seydoux, and Piero Poli.
\newblock Unsupervised learning of seismic wavefield features: clustering
  continuous array seismic data during the 2009 {L'Aquila} earthquake.
\newblock {\em Journal of Geophysical Research: Solid Earth},
  126(1):e2020JB020506, 2021.

\bibitem{Bendick-Mencin-2020:evidence}
R~Bendick and D~Mencin.
\newblock Evidence for synchronization in the global earthquake catalog.
\newblock {\em Geophysical Research Letters}, 47(15):e2020GL087129, 2020.

\bibitem{Machado-Lopes-2013:analysis}
Jos{\'e} A~Tenreiro Machado and Ant{\'o}nio~M Lopes.
\newblock Analysis and visualization of seismic data using mutual information.
\newblock {\em Entropy}, 15(9):3892--3909, 2013.

\bibitem{Ramirez-Rojas-Paez-Hernandez-Luevano-2013:cross}
A~Ram{\'\i}rez-Rojas, RT~P{\'a}ez-Hern{\'a}ndez, and J~Rub{\'e}n Lu{\'e}vano.
\newblock Cross-correlation analysis for geoelectric time series associated
  with an earthquake by means of mutual information theory.
\newblock {\em Revista Mexicana de F{\'\i}sica}, 59(1):14--17, 2013.

\bibitem{Song-Pitarka-Somerville-2009:exploring}
Seok~Goo Song, Arben Pitarka, and Paul Somerville.
\newblock Exploring spatial coherence between earthquake source parameters.
\newblock {\em Bulletin of the Seismological Society of America},
  99(4):2564--2571, 2009.

\bibitem{Ding-Trifunac-Todorovska-et-al-2015:coherence}
Haiping Ding, Mihailo~D Trifunac, Maria~I Todorovska, and Neboj{\v{s}}a
  Orbovi{\'c}.
\newblock Coherence of dispersed synthetic strong earthquake ground motion at
  small separation distances.
\newblock {\em Soil Dynamics and Earthquake Engineering}, 70:1--10, 2015.

\bibitem{Ansari-Noorzad-Zafarani-2009:clustering}
Anooshiravan Ansari, Assadollah Noorzad, and Hamid Zafarani.
\newblock Clustering analysis of the seismic catalog of {Iran}.
\newblock {\em Computers \& Geosciences}, 35(3):475--486, 2009.

\bibitem{Weatherill-Burton-2009:delineation}
Graeme Weatherill and Paul~W Burton.
\newblock Delineation of shallow seismic source zones using {K}-means cluster
  analysis, with application to the {Aegean} region.
\newblock {\em Geophysical Journal International}, 176(2):565--588, 2009.

\bibitem{Novianti-Setyorini-Rafflesia-2017:k}
Pepi Novianti, Dyah Setyorini, and Ulfasari Rafflesia.
\newblock K-means cluster analysis in earthquake epicenter clustering.
\newblock {\em International Journal of Advances in Intelligent Informatics},
  3(2):81--89, 2017.

\bibitem{Tenenbaum-Havlin-Stanley-2012:earthquake}
Joel~N Tenenbaum, Shlomo Havlin, and H~Eugene Stanley.
\newblock Earthquake networks based on similar activity patterns.
\newblock {\em Phys. Rev. E}, 86(4):046107, 2012.

\bibitem{Gozolchiani-Havlin-Yamasaki-2011:emergence}
Avi Gozolchiani, Shlomo Havlin, and Kazuko Yamasaki.
\newblock Emergence of {El} {Ni}{\~n}o as an autonomous component in the
  climate network.
\newblock {\em Phys. Rev. Lett.}, 107(14):148501, 2011.

\bibitem{Ludescher-Gozolchiani-Bogachev-et-al-2013:improved}
Josef Ludescher, Avi Gozolchiani, Mikhail~I Bogachev, Armin Bunde, Shlomo
  Havlin, and Hans~Joachim Schellnhuber.
\newblock Improved {El} {Ni}{\~n}o forecasting by cooperativity detection.
\newblock {\em Proc. Natl. Acad. Sci. U.S.A.}, 110(29):11742--11745, 2013.

\bibitem{Fan-Meng-Ashkenazy-et-al-2017:network}
Jingfang Fan, Jun Meng, Yosef Ashkenazy, Shlomo Havlin, and Hans~Joachim
  Schellnhuber.
\newblock Network analysis reveals strongly localized impacts of {El}
  {Ni}{\~n}o.
\newblock {\em Proc. Natl. Acad. Sci. U.S.A.}, 114(29):7543--7548, 2017.

\bibitem{Wang-Gozolchiani-Ashkenazy-et-al-2013:dominant}
Yang Wang, Avi Gozolchiani, Yosef Ashkenazy, Yehiel Berezin, Oded Guez, and
  Shlomo Havlin.
\newblock Dominant imprint of {Rossby} waves in the climate network.
\newblock {\em Phys. Rev. Lett.}, 111(13):138501, 2013.

\bibitem{Boers-Bookhagen-Marwan-et-al-2013:complex}
Niklas Boers, Bodo Bookhagen, Norbert Marwan, J{\"u}rgen Kurths, and Jos{\'e}
  Marengo.
\newblock Complex networks identify spatial patterns of extreme rainfall events
  of the {South} {American} {Monsoon} {System}.
\newblock {\em Geophys. Res. Lett.}, 40(16):4386--4392, 2013.

\bibitem{Quiroga-Kreuz-Grassberger-2002:event}
R~Quian Quiroga, Thomas Kreuz, and Peter Grassberger.
\newblock Event synchronization: a simple and fast method to measure
  synchronicity and time delay patterns.
\newblock {\em Phys. Rev. E}, 66(4):041904, 2002.

\bibitem{Abe-Suzuki-2003:law}
Sumiyoshi Abe and Norikazu Suzuki.
\newblock Law for the distance between successive earthquakes.
\newblock {\em J. Geophys. Res.: Solid Earth}, 108(B2), 2003.

\bibitem{Baiesi-Paczuski-2004:scale}
Marco Baiesi and Maya Paczuski.
\newblock Scale-free networks of earthquakes and aftershocks.
\newblock {\em Phys. Rev. E}, 69(6):066106, 2004.

\bibitem{Davidsen-Grassberger-Paczuski-2006:earthquake}
J{\"o}rn Davidsen, Peter Grassberger, and Maya Paczuski.
\newblock Earthquake recurrence as a record breaking process.
\newblock {\em Geophys. Res. Lett.}, 33(11), 2006.

\bibitem{Lotfi-Darooneh-2012:earthquakes}
N~Lotfi and AH~Darooneh.
\newblock The earthquakes network: the role of cell size.
\newblock {\em Euro. Phys. J. B}, 85(1):1--4, 2012.

\bibitem{Rezaei-Darooneh-Lotfi-et-al-2017:earthquakes}
Soghra Rezaei, Amir~Hossein Darooneh, Nastaran Lotfi, and Nazila Asaadi.
\newblock The earthquakes network: Retrieving the empirical seismological laws.
\newblock {\em Physica A}, 471:80--87, 2017.

\bibitem{Celikoglu-2020:earthquake}
Ahmet Celikoglu.
\newblock Earthquake spatial dynamics analysis using event synchronization
  method.
\newblock {\em Physics of the Earth and Planetary Interiors}, page 106524,
  2020.

\bibitem{He-Wang-Liu-et-al-2020:similar}
Xuan He, Luyang Wang, Zheng Liu, and Yiwen Liu.
\newblock Similar seismic activities analysis by using complex networks
  approach.
\newblock {\em Symmetry}, 12(5):778, 2020.

\bibitem{He-Wang-Zhu-et-al-2021:statistical}
Xuan He, Luyang Wang, Hongbo Zhu, and Zheng Liu.
\newblock Statistical analysis of complex weighted network for seismicity.
\newblock {\em Physica A}, 563:125468, 2021.

\bibitem{Welch-1967:use}
Peter Welch.
\newblock The use of fast {Fourier} transform for the estimation of power
  spectra: a method based on time averaging over short, modified periodograms.
\newblock {\em IEEE Transactions on audio and electroacoustics}, 15(2):70--73,
  1967.

\end{thebibliography}





\appendix
\clearpage\newpage
\section*{Supplementary Figures}

\setcounter{figure}{0}
\renewcommand{\thefigure}{S\arabic{figure}}

\begin{figure}[hbt!]
\centerline{\includegraphics[width=0.6\linewidth]{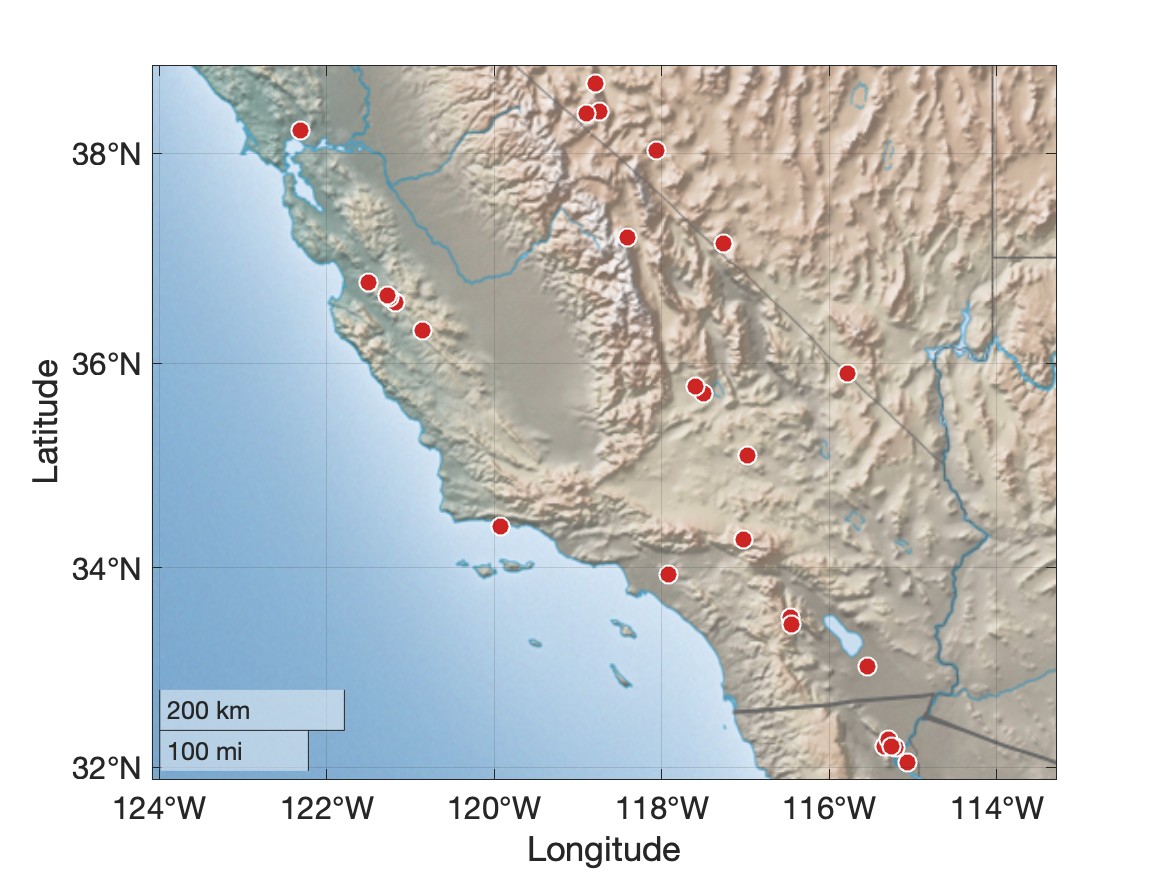}}
\caption{A map showing the locations of the 27 earthquakes (in the California region) considered in this study. The earthquakes’ magnitudes were larger than 4.5, and they occurred between April 2010 and January 2020. 
\label{fig:map_all_events}}
\end{figure}

\begin{figure}
\centerline{\includegraphics[width=\linewidth]{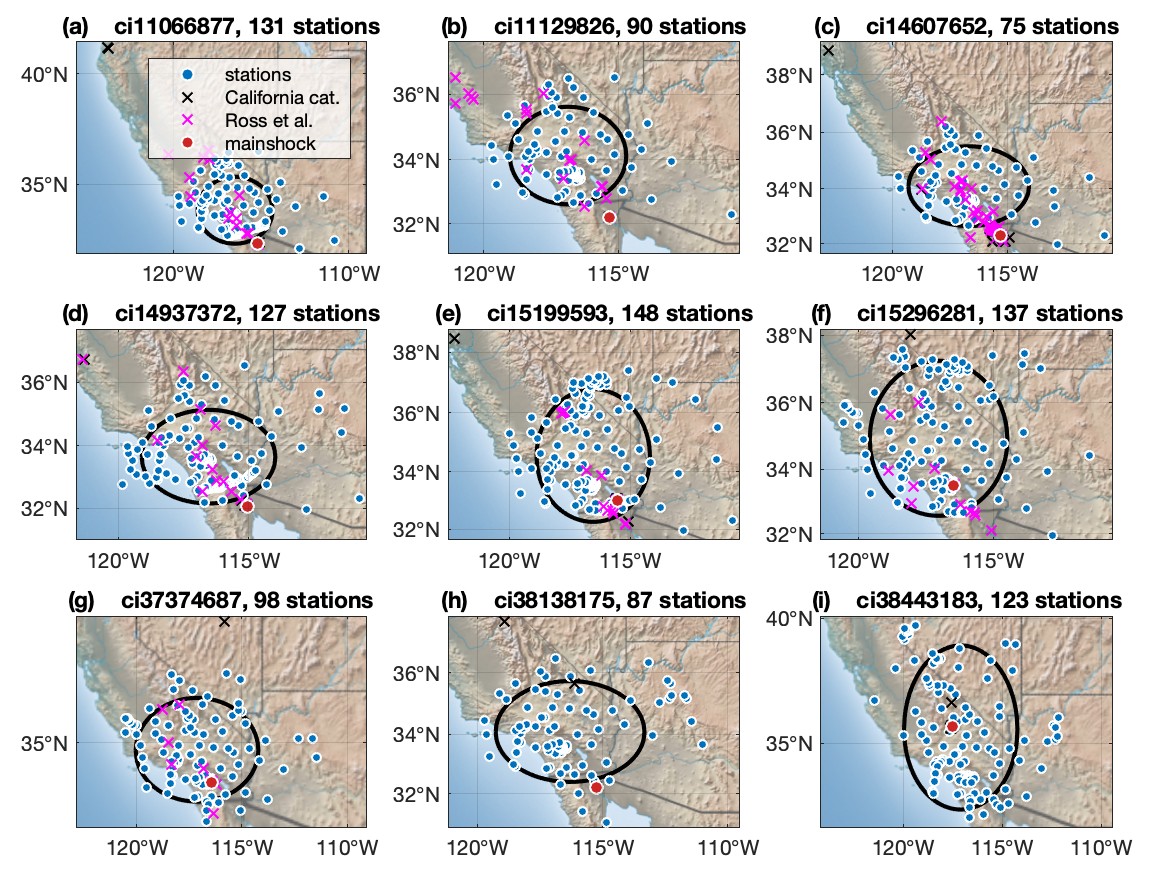}}
\caption{A map showing the location of the stations (blue) that were used for the waveform analysis, spanning 24 hours before the main earthquake (red) and one hour after it. Also included are the earthquake events of the  Ross {\it et al.} \cite{Ross-Trugman-Hauksson-et-al-2019:searching} catalog (pink “x”s) and the California catalog (black “x”s) that fall within the 25-hour window. The black ellipse indicates the region that is used to compare the network measure and the catalog events. The titles include the code of the earthquakes and the number of stations that were used in the analysis. The nine panels present nine different events. 
\label{fig:event_stations_map1}}
\end{figure}

\begin{figure}
\centerline{\includegraphics[width=\linewidth]{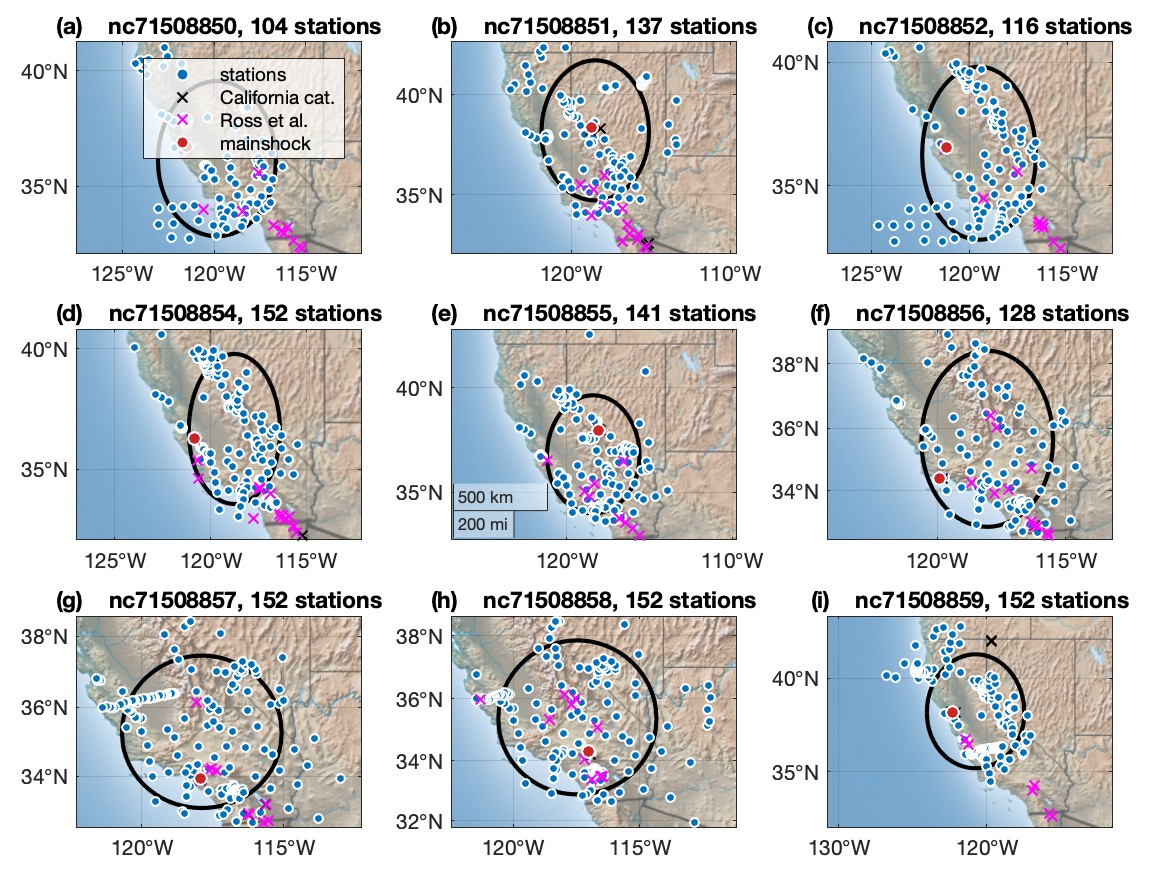}}
\caption{Same as Fig. \ref{fig:event_stations_map1} for nine different events. 
\label{fig:event_stations_map2}}
\end{figure}

\begin{figure}
\centerline{\includegraphics[width=\linewidth]{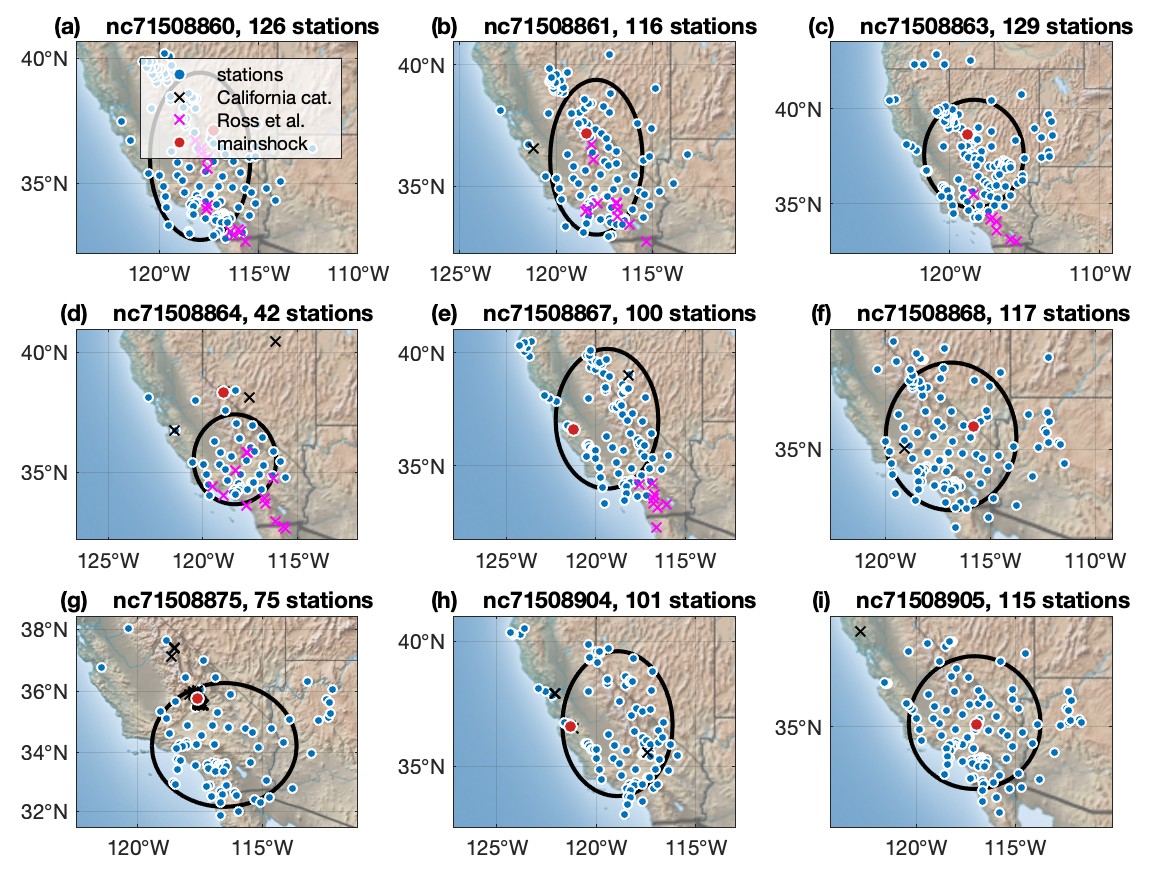}}
\caption{Same as Fig. \ref{fig:event_stations_map1} for nine different events. 
\label{fig:event_stations_map3}}
\end{figure}

\begin{figure}
\centerline{\includegraphics[width=0.9\linewidth]{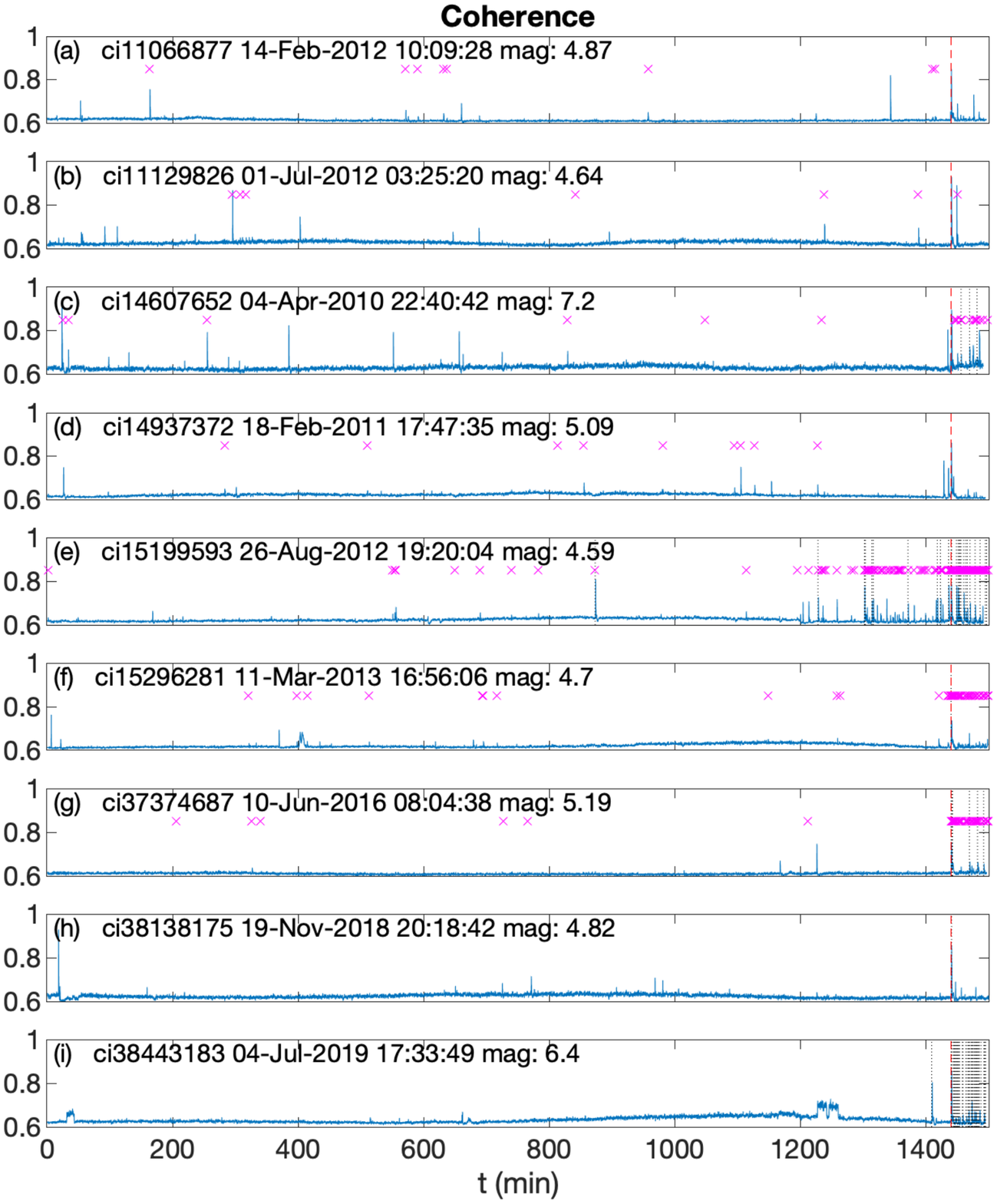}}
\caption{A 25-hour analysis of the $k$-means clustering of the coherence method. We present the analysis for events that occurred in the study region (Fig. \ref{fig:map_all_events}). We analyzed station (Fig. \ref{fig:event_stations_map1}) data starting 24 hours before the event to one hour after the event. For the clustering analysis, we used three clusters and computed the mean value within the most dominant cluster. Each panel in the figure depicts the time series corresponding to different events. The dashed vertical red line indicates the mainshock, the dotted vertical lines indicate the other events included in the California catalog, and the pink $\times$s indicate the earthquakes included in the  Ross {\it et al.} \cite{Ross-Trugman-Hauksson-et-al-2019:searching} catalog and that are larger than 1.3. Shown is the analysis of nine different events of Fig. \ref{fig:event_stations_map1}. Each panel includes the code, time, and magnitude of the event.
\label{fig:cluster1}}
\end{figure}

\begin{figure}
\centerline{\includegraphics[width=\linewidth]{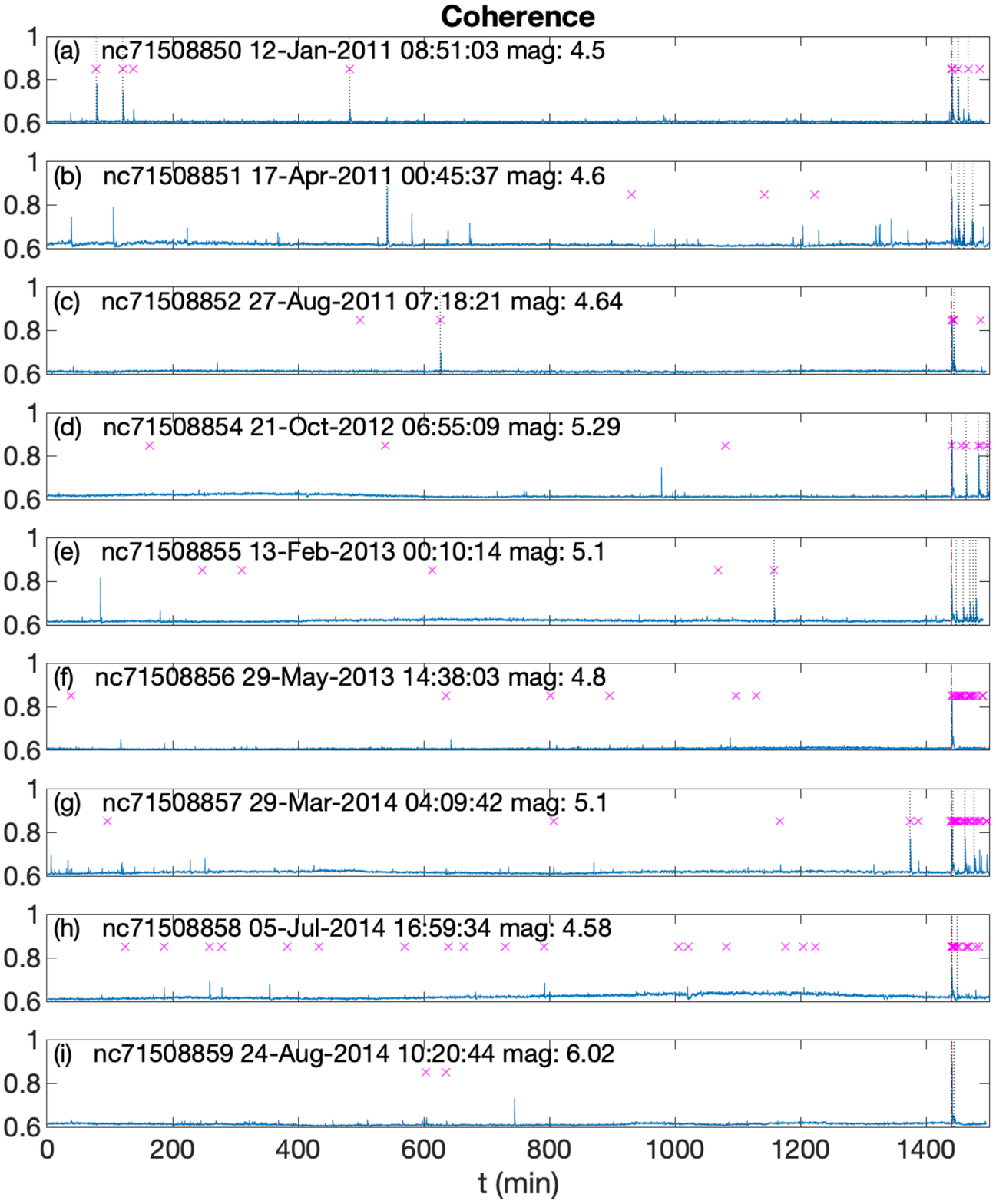}}
\caption{Same Fig. \ref{fig:cluster1} for the nine different events of Fig. \ref{fig:event_stations_map2}. 
\label{fig:cluster2}}
\end{figure}

\begin{figure}
\centerline{\includegraphics[width=\linewidth]{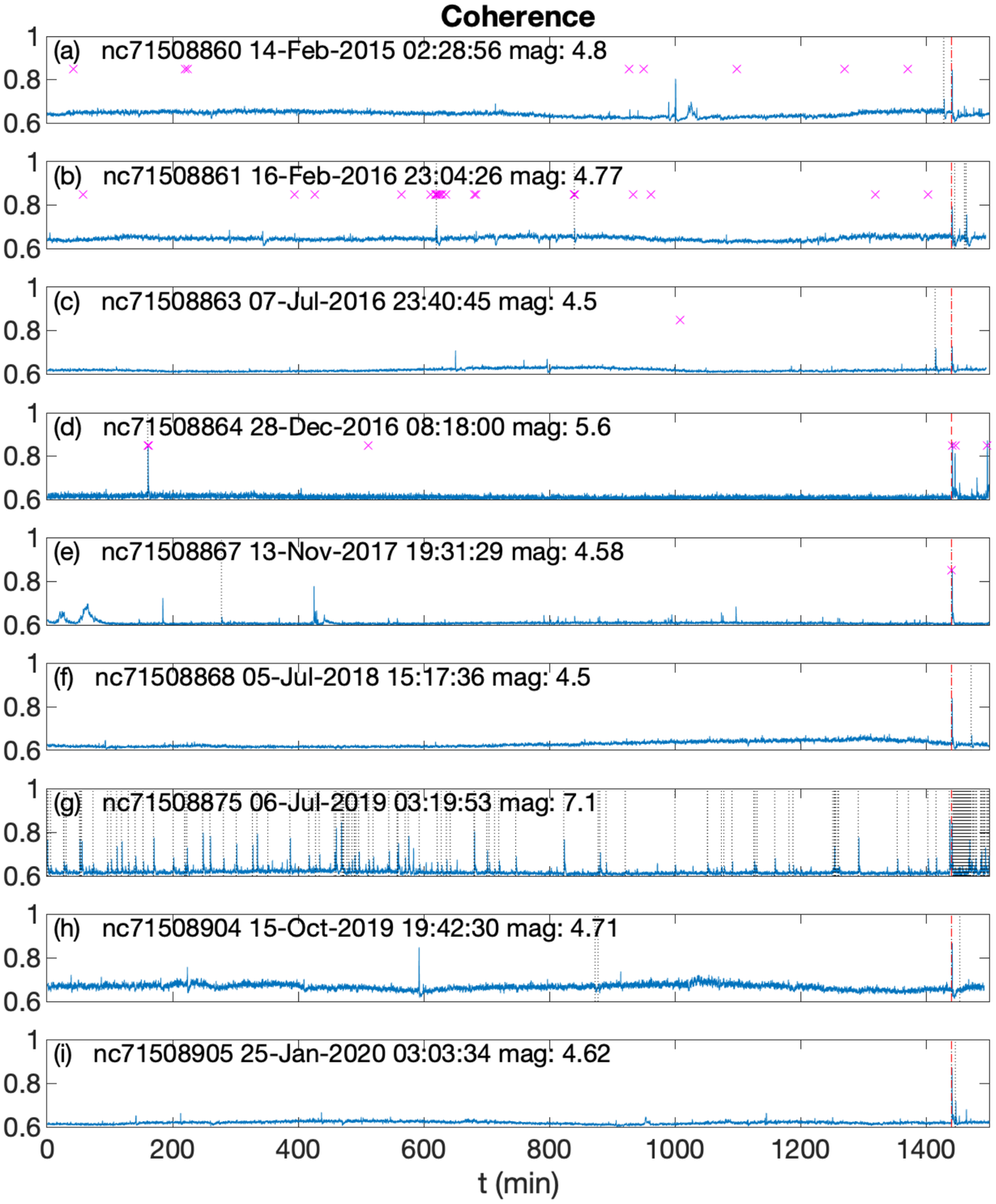}}
\caption{Same Fig. \ref{fig:cluster1} for the nine different events of Fig. \ref{fig:event_stations_map3}.
\label{fig:cluster3}}
\end{figure}

\end{document}